\documentclass{iopart}

\usepackage{graphicx}
\usepackage{dcolumn}
\usepackage{bm}
\usepackage{ifpdf}
\usepackage{epstopdf}
\usepackage{color}
\usepackage{CJK}
\usepackage{amssymb, iopams}

\usepackage{ulem}

\newcommand{\delete}{\bgroup\markoverwith{\textcolor{red}{\rule[0.5ex]{2pt}{1pt}}}\ULon}

\begin{document}

\title{Covariant density functional analysis of shape evolution in $N =40$ isotones}

\author{Z. H. Wang$^1$, J. Xiang$^1$, W. H. Long$^1$ and Z. P. Li$^2$}
\address{$^1$ School of Nuclear Science and Technology, Lanzhou University, Lanzhou 730000, China} \ead{longwh@lzu.edu.cn}
\address{$^2$ School of Physical Science and Technology, Southwest University, Chongqing 400715, China}\ead{zpliphy@swu.edu.cn}

\date{\today}

\begin{abstract}
  The structure of low-lying excitation states of even-even $N=40$ isotones is studied using a five-dimensional collective Hamiltonian with the collective parameters determined from the relativistic mean-field plus BCS method with the PC-PK1 functional in the particle-hole channel and a separable paring force in the particle-particle channel. The theoretical calculations can reproduce not only the systematics of the low-lying states along the isotonic chain but also the detailed structure of the spectroscopy in a single nucleus. We find a picture of spherical-oblate-prolate shape transition along the  isotonic chain of $N=40$ by analyzing the potential energy surfaces. The coexistence of low-lying excited $0^+$ states has also been shown to be a common feature in neutron-deficient $N=40$ isotones.
\end{abstract}

\maketitle

\setlength{\abovecaptionskip}{0em}

\section{\label{Introduction}Introduction}

Shapes of exotic nuclei far from stability have been extensively explored in many experimental and theoretical studies. The evolution of ground-state shapes along an isotopic or isotonic chain is governed by changes of the underlying shell structure of single-nucleon levels. In particular far from the $\beta$-stability line, the energy spacings between single-nucleon levels change considerably with respect to the neutron or proton numbers. This can lead to reduced spherical shell/subshell gaps~\cite{Sorlin2008}, and further result in the onset of ground-state deformation or the coexistence of different shapes in a single nucleus.

In the last decades the evolution of shapes and underlying shell structure in the $N=40$ isotones have attracted lots of attentions. In a simplistic shell-model~\cite{mayer1955}, $N=40$ is predicted to be a subshell since neutrons fill up the $fp$ shell and leave the orbit $1g_{9/2}$ remaining empty. This is supported by the measurements of the low-lying excited states and transition rates in $^{68}$Ni, namely a rather high $2^+_1$ state, the first excited $0^+$ lower than $2^+_1$ state~\cite{Bernas1982279} and very small $B(E2; 2^+_1\to 0^+_1)$~\cite{PhysRevLett.88.092501}. However, when removing protons from $^{68}$Ni several experiments of $\beta$ decay suggest a rapidly weakening of $N = 40$ gap~\cite{PhysRevLett.83.3613,Sorlin2001183}, and result in the enhanced quadrupole collectivity in the neutron-rich $N=40$ isotones. This is  confirmed by the recent measurements of the first excited $2^+$ states in $^{64}$Cr and $^{66}$Fe~\cite{PhysRevC.81.051304}. Furthermore, Tarasov and co-workers suggested a new island of inversion in the region of $^{62}$Ti from systematics of production cross sections~\cite{Tarasov2009}. Toward the neutron deficient side, there exists the coexistence of complex shapes in some nuclei, such as the light Kr and Se isotopes~\cite{Ljungvall2008,PhysRevC.75.054313}.

On the theoretical side, a number of models have been used to discuss the emergency of collectivity in  $N=40$ isotones. A shell-model-based calculation shows that the size of $N=40$ gap of $^{68}$Ni is not large enough to prevent paring correlation from developing~\cite{PhysRevLett.88.092501}. Furthermore, it is found in the interacting shell-model calculations with a large valence  space~\cite{Lenzi2010}  that the onset of deformation emerges at $N=40$ in the iron chain and at $N=38$ in chromium isotopes. The collective structure of $N=40$ isotones have been also studied by five-dimensional collective Hamiltonian (5DCH) based on the Harteee-Fock-Bogoliubov (HFB) model with the Gogny force D1S~\cite{PhysRevC.80.064313}. It was predicted that the collectivity grows with $Z$ and  shape coexistence occurs in the neutron-deficient $N=40$ isotones above$^{74}$Se.

{Since 1980s, the covariant density functional theory (CDFT) \cite{Walecka:1974, Serot:1986} becomes more and more popular in exploring the ground state and excitation properties of both spherical and deformed nuclei all over the nuclear chart. One of the typical candidates is the} relativistic mean-field (RMF) theory~\cite{reinhard1989relativistic, Ring1996193, vretenar2005relativistic, meng2006relativistic} based on the finite-range meson exchange picture \cite{Yukawa:1935}. Recently under the limit of zero range, the CDFT with a point-coupling interaction has attracted more and more attentions~\cite{Niksic2011} because of its advantages in the extensions for nuclear low-lying excited states by implementing projection techniques~\cite{jiang2008three,PhysRevC.79.044312} , the generator coordinate methods \cite{PhysRevC.73.034308, PhysRevC.74.064309, PhysRevC.81.044311, PhysRevC.83.014308, PhysRevC.84.024306} and collective Hamiltonian~\cite{PhysRevC.79.034303}. Specifically, the CDFT-based 5DCH has achieved great success in describing the low-lying states of finite nuclei ranging from $A\sim40$ to superheavy regions, including the spherical, transitional, and well-deformed  ones~\cite{Niksic2011,PhysRevC.82.054319, PhysRevC.79.054301, LiZhipan2010, PhysRevC.81.034316, PhysRevC.81.034311, LiZhipan2009, PhysRevC.79.034303, LiZhipan2011, MeiHua2012, LiZhipan2012PLB}.

In this work, the shape evolution and low-lying states  will be studied along the isotonic chain of $N=40$ by the 5DCH based on CDFT, and the covariant density functional PC-PK1~\cite{PhysRevC.82.054319} is used in the particle-hole channel whereas in the particle-particle channel a separable pairing force \cite{Tian200944} is taken as the effective interaction. The content is arranged as follows. In Sec.\ref{framework }, we will give a brief introduction for the 5DCH based on CDFT with PC-PK1. In Sec.\ref{results}, the shape evolution of even-even nuclei along $N=40$ isotonic chain and shape coexistence in $^{76}$Kr, $^{78}$Sr, and $^{80}$Zr will be discussed. In the end, a summary is presented in Sec.\ref{summary}.

%
\section{\label{framework }Theoretical framework  }
%

The formalisms of CDFT  based on point-coupling density functionals have been elaborated in a variety of literatures~\cite{Ring1996193,vretenar2005relativistic,PhysRevC.79.034303, meng2006relativistic,xiang2012covariant} and will not be repeated here.

The quantized 5DCH that describes the nuclear rotational excitations, quadrupole vibrations, and their couplings can be written as the following form~\cite{PhysRevC.79.054301, prochniak2004self, PhysRevC.79.034303}
\begin{equation}
\hat H_{\rm coll}=\hat T_{\rm rot}(\beta,\gamma,\Omega)+\hat T_{\rm vib}(\beta,\gamma)+V_{\rm coll}(\beta,\gamma),
\label{Hami}
\end{equation}
where $V_{\rm coll}$ is the collective potential, and $\hat T_{\rm rot}$ and $\hat T_{\rm vib}$ are respectively the rotational  and vibrational kinetic energies,
\begin{eqnarray} \fl
\hat T_{\rm rot}  & =&  \frac{1}{2}\sum_{k=1}^3\frac{\hat{J_k}^2}{\mathcal I_k}, \label{Trot}\\[0.25em] \fl
\hat T_{\rm vib}  &=&  -\frac{\hbar^2}{2\sqrt{\omega\gamma}}\Big\{\frac{1}{\beta^4} \Big[ \frac{\partial}{\partial\beta}\sqrt{\frac{\gamma}{\omega}}\beta^4 B_{\gamma\gamma}\frac{\partial}{\partial\beta}  - \frac{\partial}{\partial\beta}\sqrt{\frac{\gamma}{\omega}}\beta^3 B_{\beta\gamma}\frac{\partial}{\partial\gamma}\Big]+\nonumber\\ \fl
&&\hspace{4.75em}\frac{1}{\beta\sin3\gamma}\Big[-\frac{\partial}{\partial\gamma} \label{Tvib} \sqrt{\frac{\gamma}{\omega}}\sin3\gamma B_{\beta\gamma} \frac{\partial}{\partial\beta}+\frac{1}{\beta}\frac{\partial}{\partial\gamma} \sqrt{\frac{\gamma}{\omega}}\sin3\gamma B_{\beta\beta}\frac{\partial}{\partial\gamma}\Big]\Big\}, \\[0.25em] \fl
V_{\rm coll}         & = & E_{\rm tot}-\Delta V_{\rm vib}-\Delta V_{\rm rot}.\label{vcoll}
\end{eqnarray}
In above expressions, ${\hat{J}}_k$ denotes the $k$ component of the angular momentum in the body-fixed frame of a nucleus, and the mass parameters $B_{\beta\beta}$, $B_{\beta\gamma}$, $B_{\gamma\gamma}$ and the moment of inertia  $\mathcal{I}_k$ depend on the quadrupole deformation variables $\beta$ and $\gamma$. For the moment of inertia $\mathcal I_k$, it can be expressed as,
\begin{eqnarray}
\mathcal{I}_k =& 4B_k\beta^2\sin^2(\gamma-2k\pi/3),\quad\quad k= 1, 2, 3,
\end{eqnarray}
where $B_k$ denotes the inertia parameter. In Eq. (\ref{Tvib}), the variables $r=B_1B_2B_3$ and $w=B_{\beta\beta}B_{\gamma\gamma}-B_{\beta\gamma}^2 $ define the volume element of the collective space. In Eq.~(\ref{vcoll}) $E_{\rm tot}$ is the binding energy determined by the constrained CDFT calculations. The $\Delta V_{\rm vib}$ and $\Delta V_{\rm rot}$ are the zero-point-energies (ZPE) of the vibrational and rotational motions, respectively~\cite{Girod197940, PhysRevC.79.054301}. The corresponding eigenvalue problem is solved by expanding the eigenfunctions on a complete set of basis functions in the collective space of the quadrupole deformations $(\beta, \gamma)$ and Euler angles $(\Omega =\phi, \theta, \psi)$.

The dynamics of the 5DCH is governed by seven functions of the intrinsic deformations $\beta$ and $\gamma$: the collective potential $V_{\rm coll}$, three mass parameters $B_{\beta\beta}$, $B_{\beta\gamma}$, $B_{\gamma\gamma}$, and three moments of inertia $\mathcal{I}_k$. These functions are determined using the cranking approximation formula based on the intrinsic triaxially deformed mean-field states, generated by the constrained CDFT calculations with respect to the intrinsic deformations $(\beta, \gamma)$. The diagonalization of the Hamiltonian~(\ref{Hami}) yields the excitation energies and collective wave functions which are used to calculate the observables~\cite{PhysRevC.79.034303}.

In the CDFT calculation, parity ($\hat P$), x-simplex symmetry ($\hat P e^{i\pi\hat J_x}$), and time-reversal invariance are imposed for the single-particle states. The Dirac equation is solved by expanding the spinors on the basis of eigenfunctions of a three-dimensional harmonic oscillator in Cartesian coordinates with 12 major shells, which are found to be sufficient to obtain a reasonably convergent results for the nuclei in this mass region. The numerical details of the calculations can also be found in Refs.~\cite{PhysRevC.79.034303,xiang2012covariant}.

%
\section{RESULTS AND DISCUSSION \label{results}}
%

\subsection{Shape evolution of $N=40$ isotones}
\begin{figure}[htbp]

\begin{flushright}
\ifpdf
\includegraphics[width=0.8\textwidth]{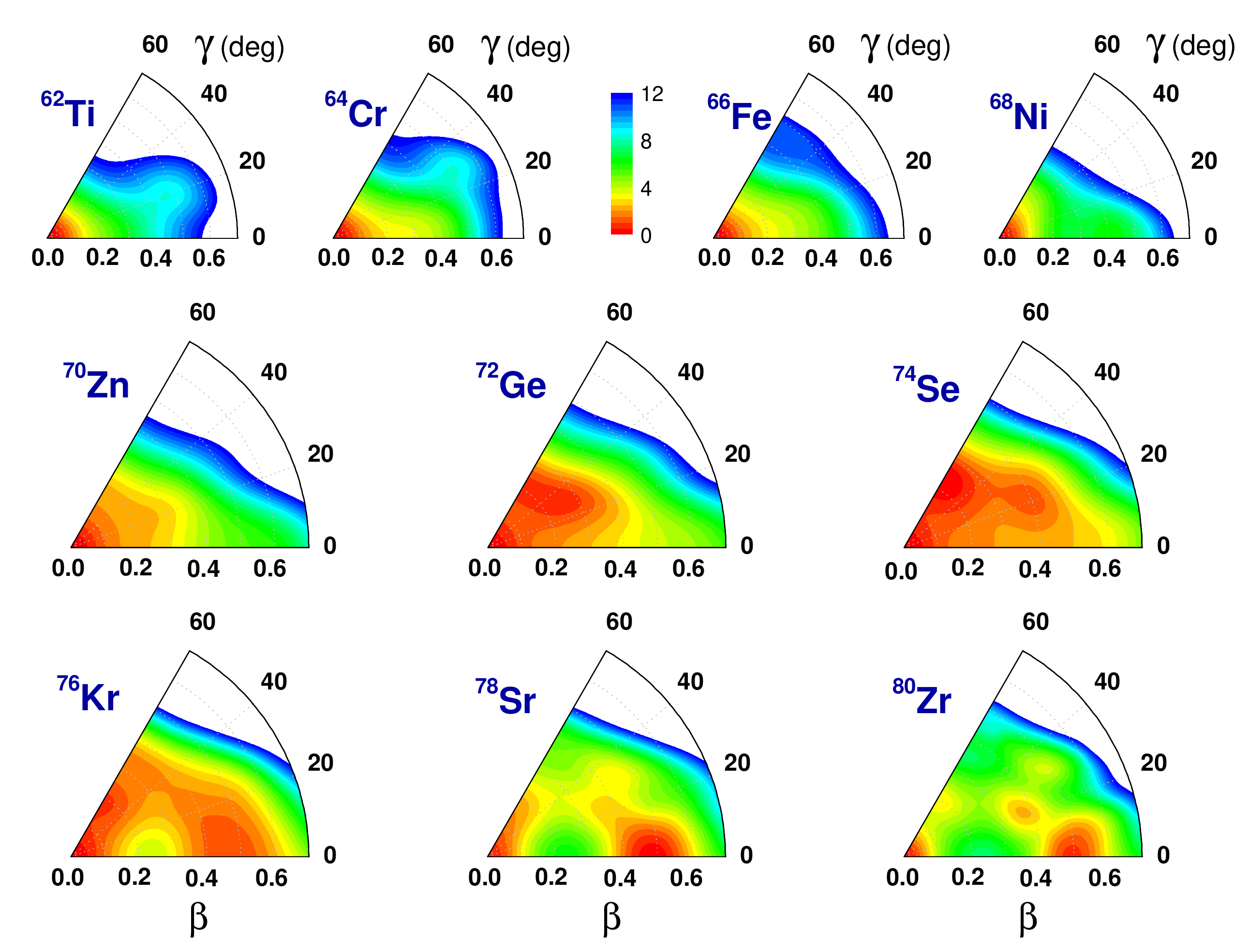}
\else
\includegraphics[width=0.8\textwidth]{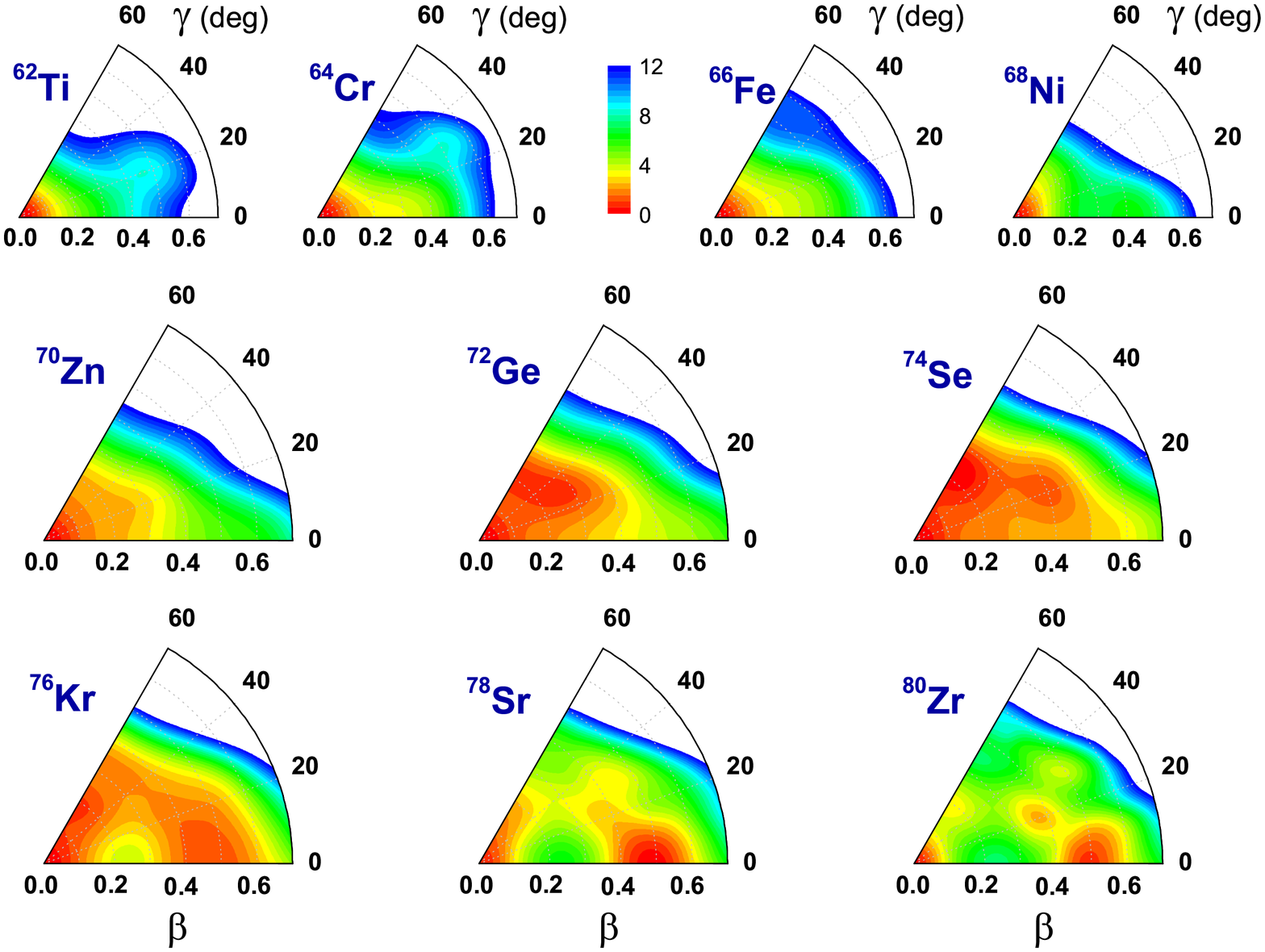}
\fi
\end{flushright}
\caption{(Color online) Potential energy surfaces of even-even $N=40$ isotones in the $\beta-\gamma$ plane extracted from the constrained RMF calculations using the PC-PK1 functional plus BCS pairing. All the energies are normalized to the absolute minimum. The energy difference between neighboring contour lines is 0.5~MeV. See the text for details.}
\label{fig:PES}
\end{figure}

Figure \ref{fig:PES} displays the potential energy surfaces (PESs) of even-even $N=40$ isotones from $^{62}$Ti to $^{80}$Zr. Similar results have been obtained by the HFB calculations with the Gogny force D1S~\cite{PhysRevC.80.064313}. The PESs nicely illustrate the rapid shape evolution along the isotonic chain. Below $^{68}$Ni the nuclei are spherical, and $^{70}$Zn becomes rather soft along the $\beta$ direction although the global minimum locates at spherical point. When adding more protons, a triaxial minimum  at ($\beta\approx0.25, \gamma\approx36^\circ$) emerges in $^{72}$Ge, whereas in $^{74}$Se  an extended oblate minimum is found. Further the coexistence of an extended oblate minimum and a large deformed prolate minimum occurs in $^{76}$Kr and these two minima are connected by a rather low triaxial barrier. Extending to $Z=38$ and $40$, the prolate minima become deeper and the triaxial barriers between the prolate and spherical minima rise higher.

\begin{figure}[htbp]
\begin{flushright}
\ifpdf
\includegraphics[width=0.8\textwidth]{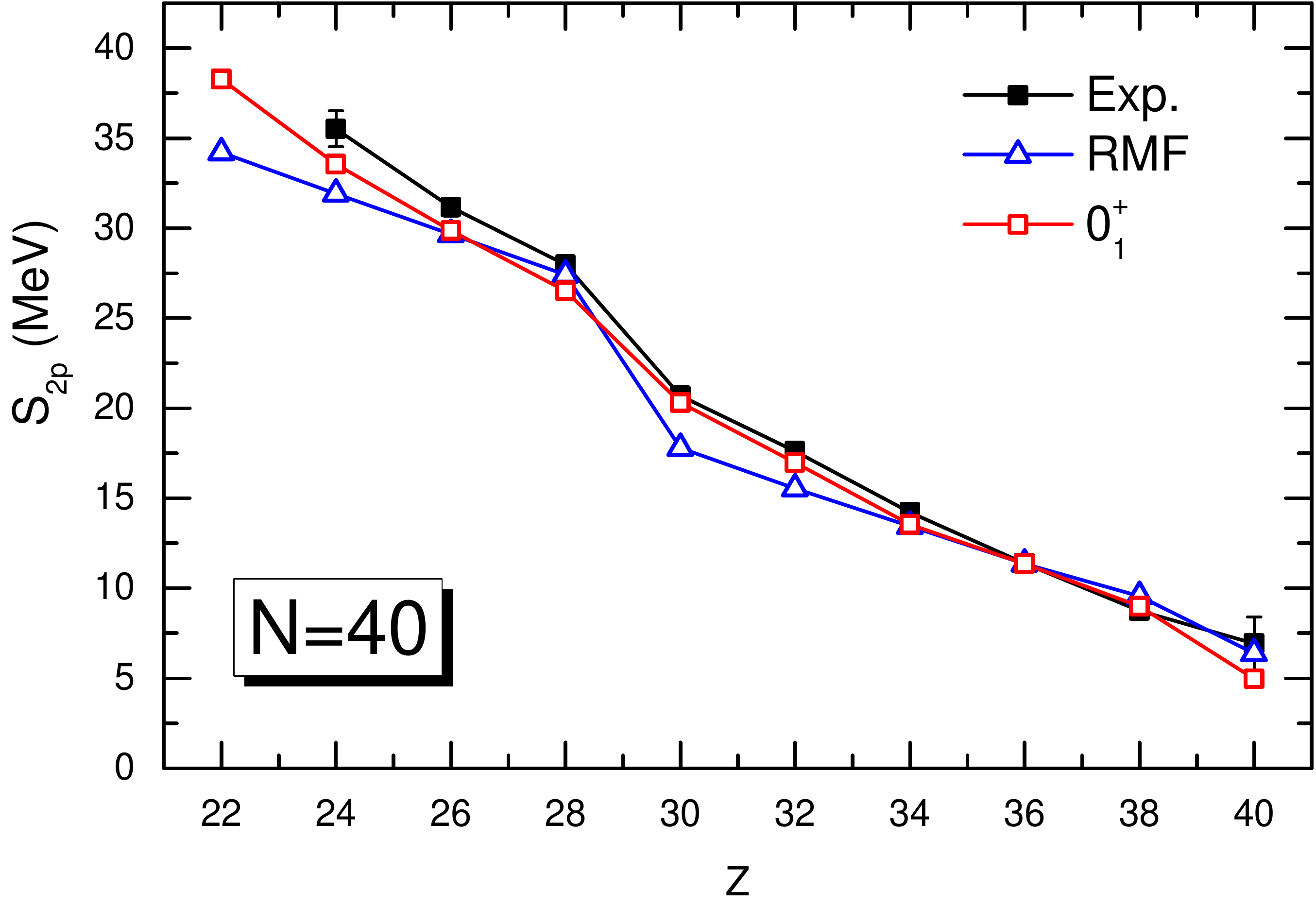}
\else
\includegraphics[width=0.8\textwidth]{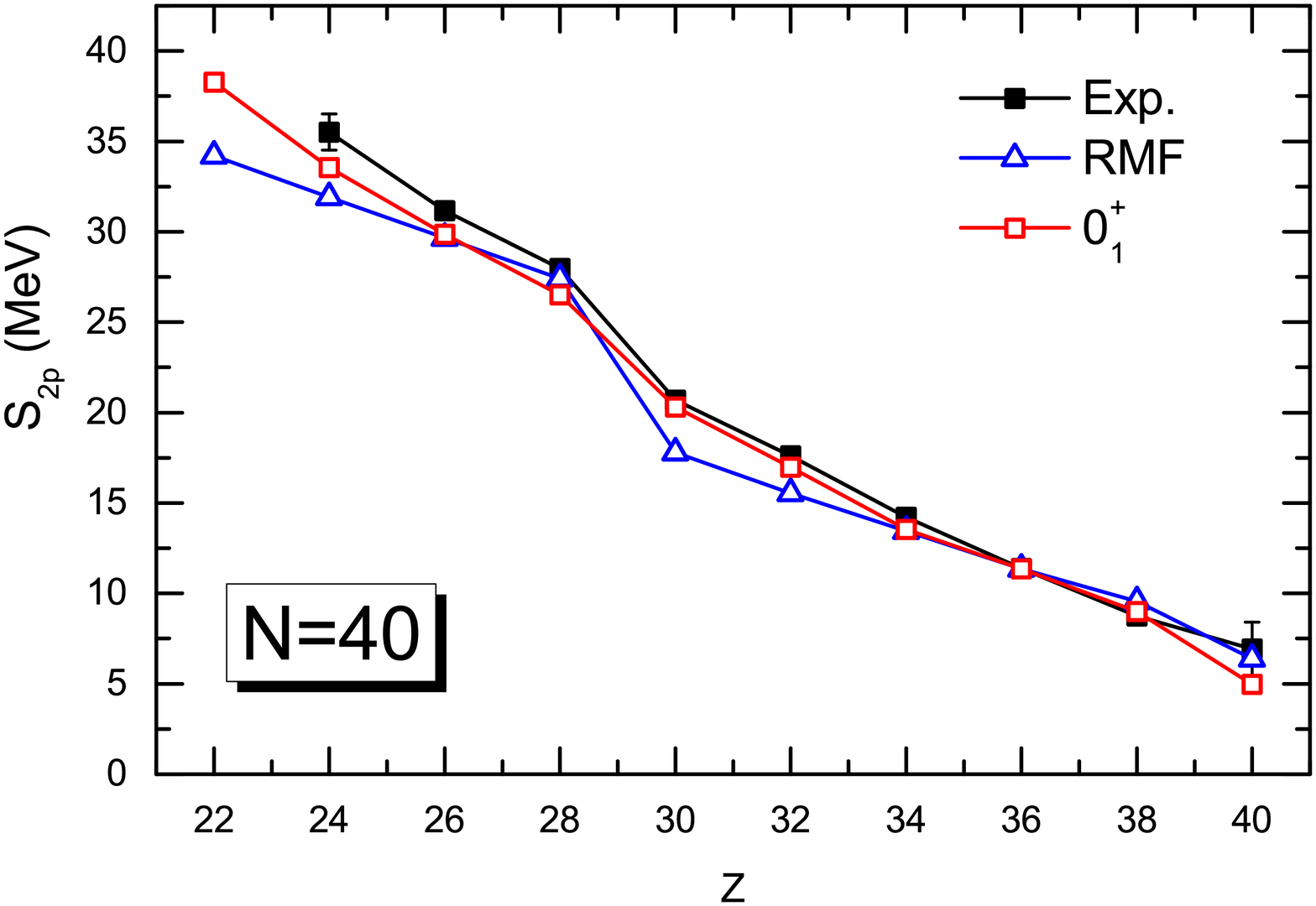}
\fi
\end{flushright}
\caption{(Color online) The evolution of two-proton separation energies $S_{2p}$ along the isotonic chain of $N=40$. The theoretical values of $S_{2p}$ extracted from the binding energies of the global minima (open triangles) and the energies of ground states given by 5DCH (open squares) are compared with the data (solid squares). See the text for details.}
\label{fig:S2p}
\end{figure}

The evolution of two-proton separation energies for $N=40$ isotones is shown in Fig.~\ref{fig:S2p}, where the theoretical results are extracted from the binding energies of the global minima (in open triangles) and the energies of ground states given by 5DCH (in open squares), as compared to the data \cite{audi2012nubase2012}. The calculations reproduce the data quite well. In particular, when the dynamical correlation energies are included by 5DCH, the root-mean-square (rms) deviation from the experiment data is reduced from 1.82~MeV to 1.18~MeV significantly.

\begin{figure}[htbp]
\begin{flushright}
\ifpdf
\includegraphics[width=0.9\textwidth]{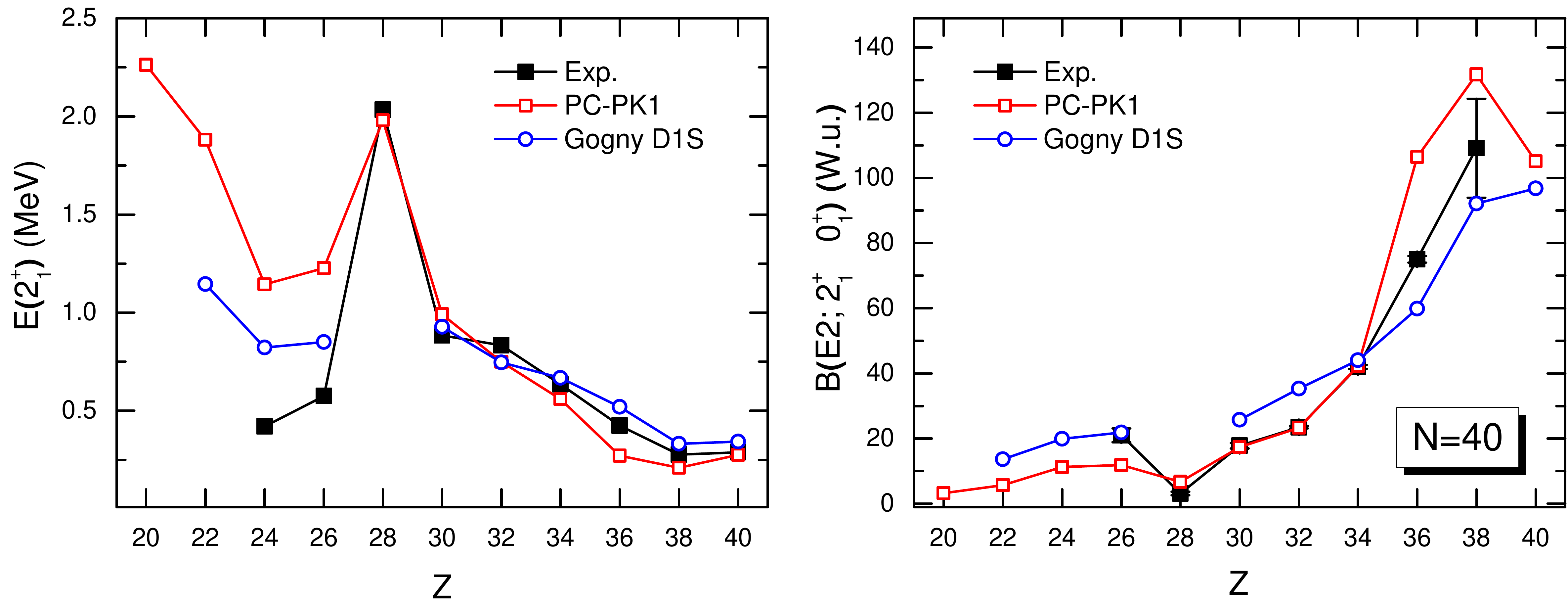}
\else
\includegraphics[width=0.9\textwidth]{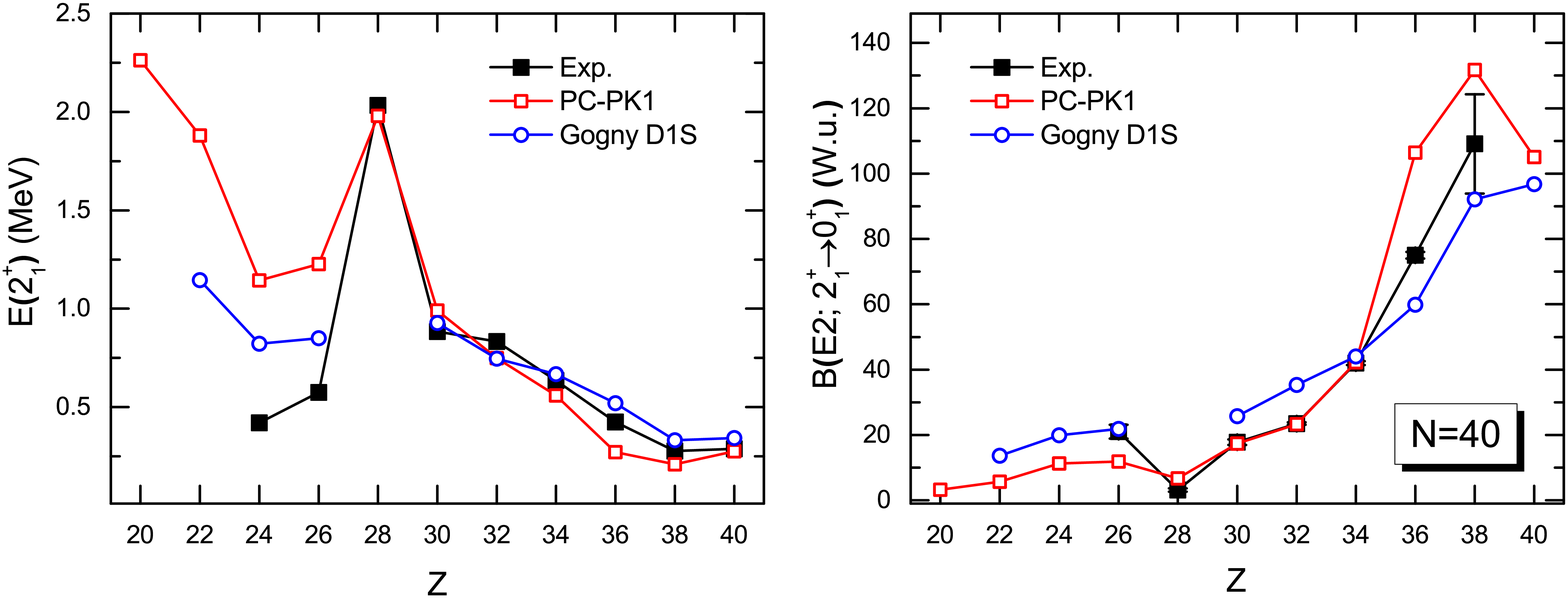}
\fi
\end{flushright}
\caption{(Color online) Excitation energies of $2_1^+$ states and transition probabilities $B(E2;2^+_1\rightarrow0^+_1)$ as a function of proton number for the $N=40$ isotones, in comparison with the similar calculations using the Gogny force D1S~\cite{PhysRevC.80.064313} and the available data~\cite{Brookhaven}. See the text for details.}
\label{fig:E2BE2}
\end{figure}

Two of the distinct features for shape transitions are the excitation energy of the first $2^+$ state and the reduced $E2$ transition probability $B(E2;2^+_1\rightarrow0^+_1)$, which are shown in  Fig.~\ref{fig:E2BE2} for the $N=40$ isotones. In Ref.~\cite{PhysRevC.80.064313}, similar calculations have been done using the Gogny force D1S and it has been also adopted in a global study of the low-lying states of the nuclei with proton numbers ranging from $Z = 10$ to $Z = 110$ and neutron numbers $N\leqslant 200$~\cite{Delaroche2010}. For comparison, the corresponding results are plotted in Fig. \ref{fig:E2BE2}. It can be seen that the 5DCH calculations based on the PC-PK1 functional give results similar to those obtained by the Gogny force D1S and both reproduce the systematics of the low-lying states for the isotones with $Z>28$, particularly for the onset of the enhanced collectivity around $Z=38$. For the neutron rich side, both calculations overestimate the excitation energies of $2_1^+$ states,leading to underestimated collectivity.

\begin{figure}[!htbp]
\begin{flushright}
\ifpdf
\includegraphics[width=0.8\textwidth]{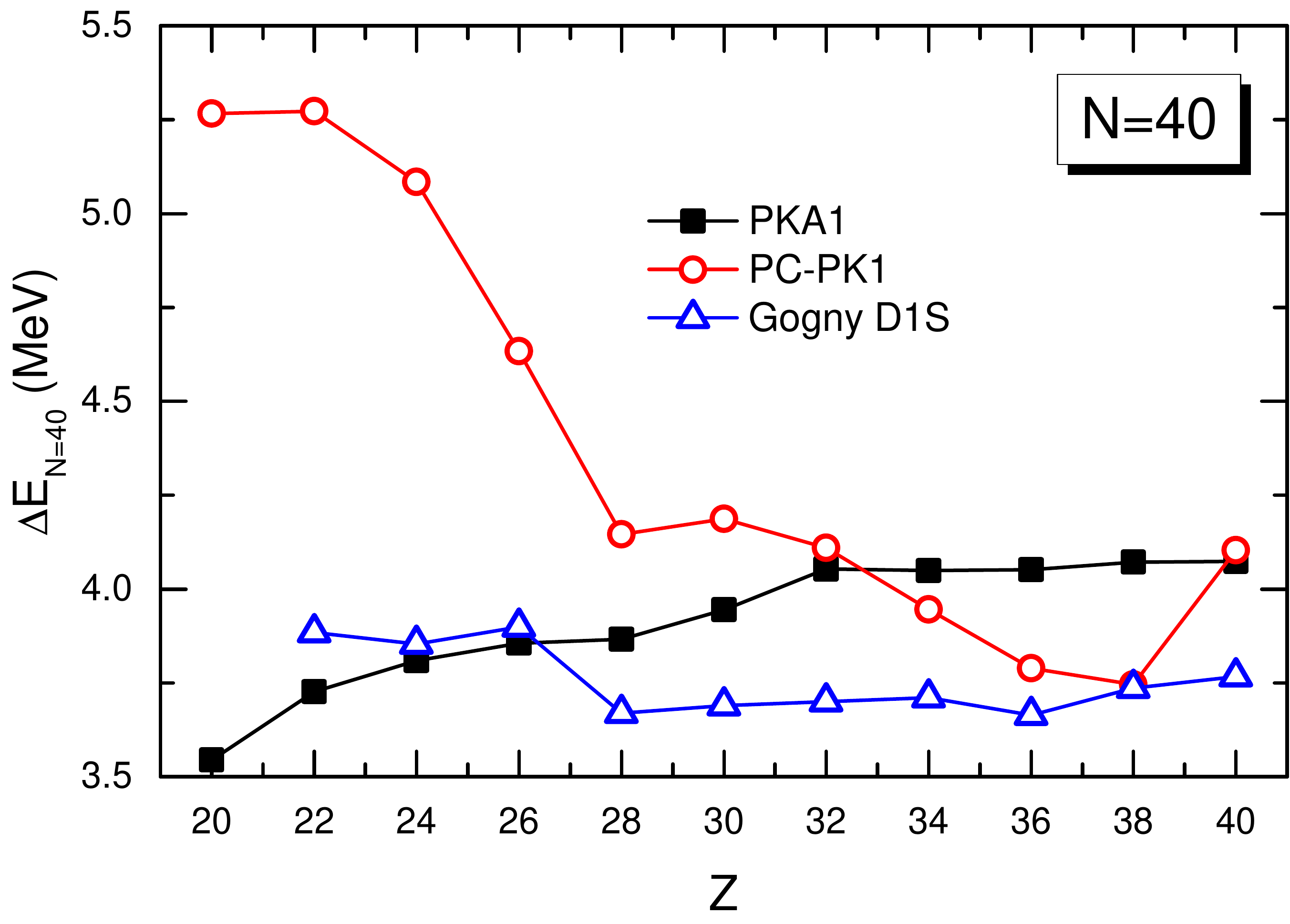}
\else
\includegraphics[width=0.8\textwidth]{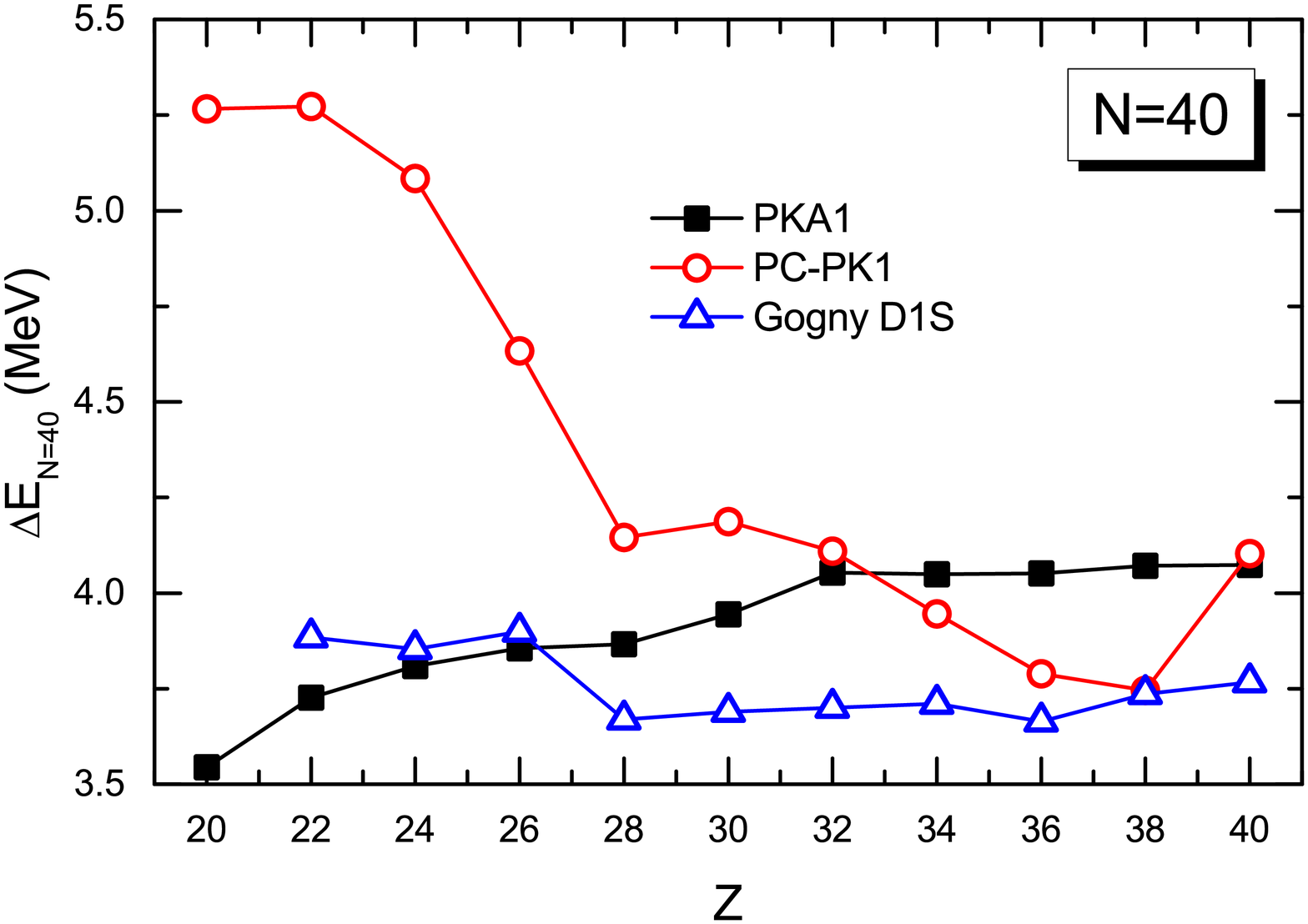}
\fi
\end{flushright}
\caption{(Color online) Evolution of the spherical neutron shell gaps at $N=40$ along the isotonic chain of $N=40$. The results are calculated by RMF+BCS with PC-PK1 parameter set,  as compared to those by DDRHF+BCS with PKA1 effective interaction and by HFB with Gogny force D1S. See the text for details.}
\label{fig:delta}

\end{figure}

To understand the lost collectivity in the neutron rich side, in Fig.~\ref{fig:delta} we plot the evolution of $N=40$ spherical neutron shell gaps since these nuclei have the spherical global minima (see Fig. \ref{fig:PES}). The calculations are performed  by RMF+BCS with PC-PK1 parameter set and the density-dependent relativistic Hartree-Fock theory \cite{Long:2006} plus BCS pairing (DDRHF+BCS) with the effective interaction PKA1~\cite{Long2007}. Moreover, the results calculated by HFB using Gogny force D1S are also shown for comparison. Generally, the enhanced spherical shell gap indicates a weakening collectivity, and vise versa. In Fig.~\ref{fig:delta} the $N=40$ shell gaps predicted by PC-PK1 and D1S increase when removing protons from $^{68}$Ni, and result in too high $2^+_1$ states. In contrast, the shell gaps given by PKA1 decrease gradually as the proton number decreases, consistent with the conclusion given by the experiments of $\beta$-decay~\cite{PhysRevLett.83.3613,Sorlin2001183}. This is mainly due to the fact that PKA1 contains the tensor components in the Fock terms of $\pi$ pseudo-vector and $\rho$ tensor couplings~\cite{Long2008, Wang2013, Jiang:2014}. The work on enhanced collectivity of $N=40$ neutron-rich isotones using 5DCH based on deformed DDRHF+BCS with PKA1 is in progress.

\subsection{Shape coexistence in $^{78}$Sr and $^{80}$Zr}

\begin{figure}[!htbp]
\begin{flushright}
\ifpdf
\includegraphics[width=0.8\textwidth]{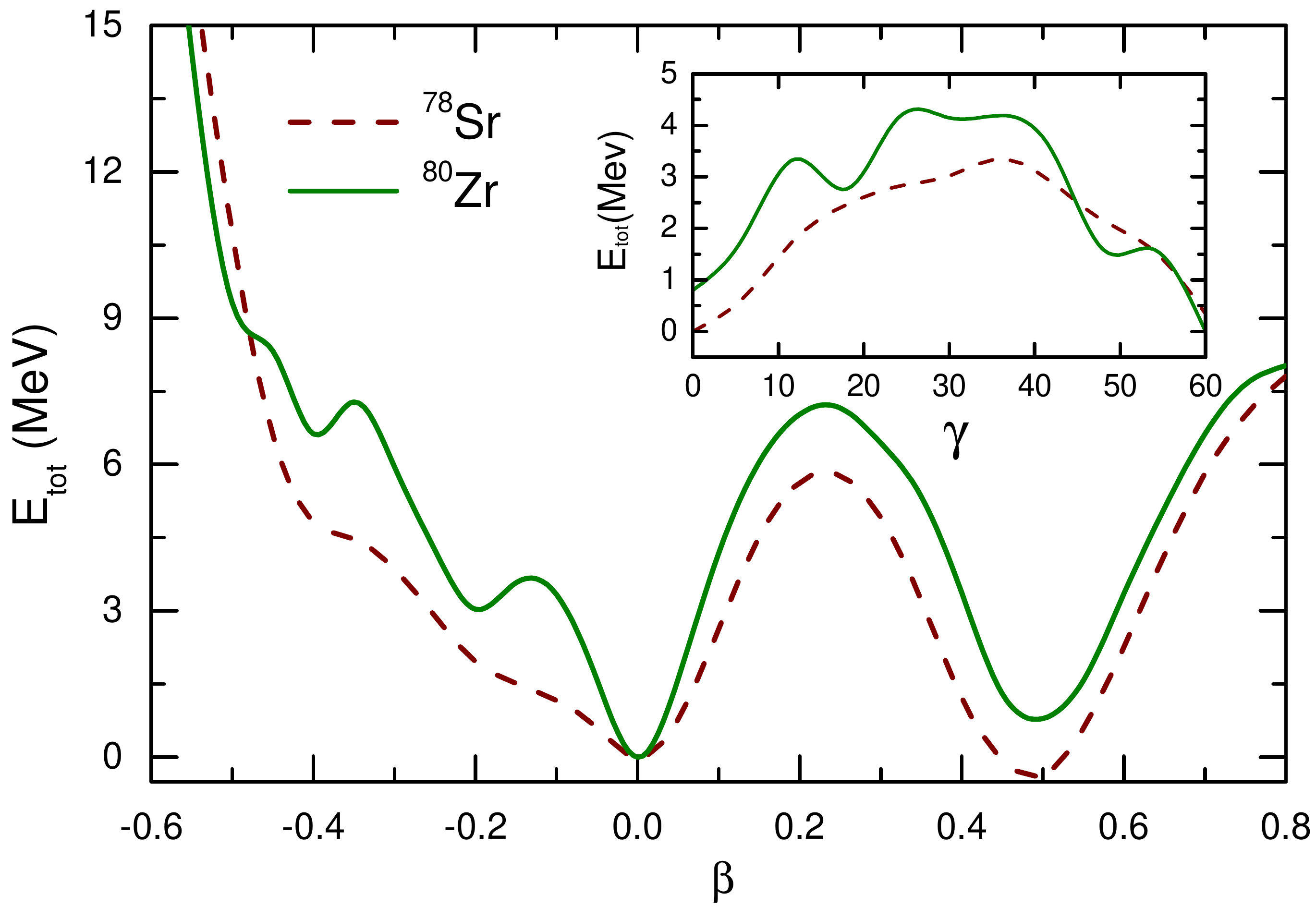}
\else
\includegraphics[width=0.8\textwidth]{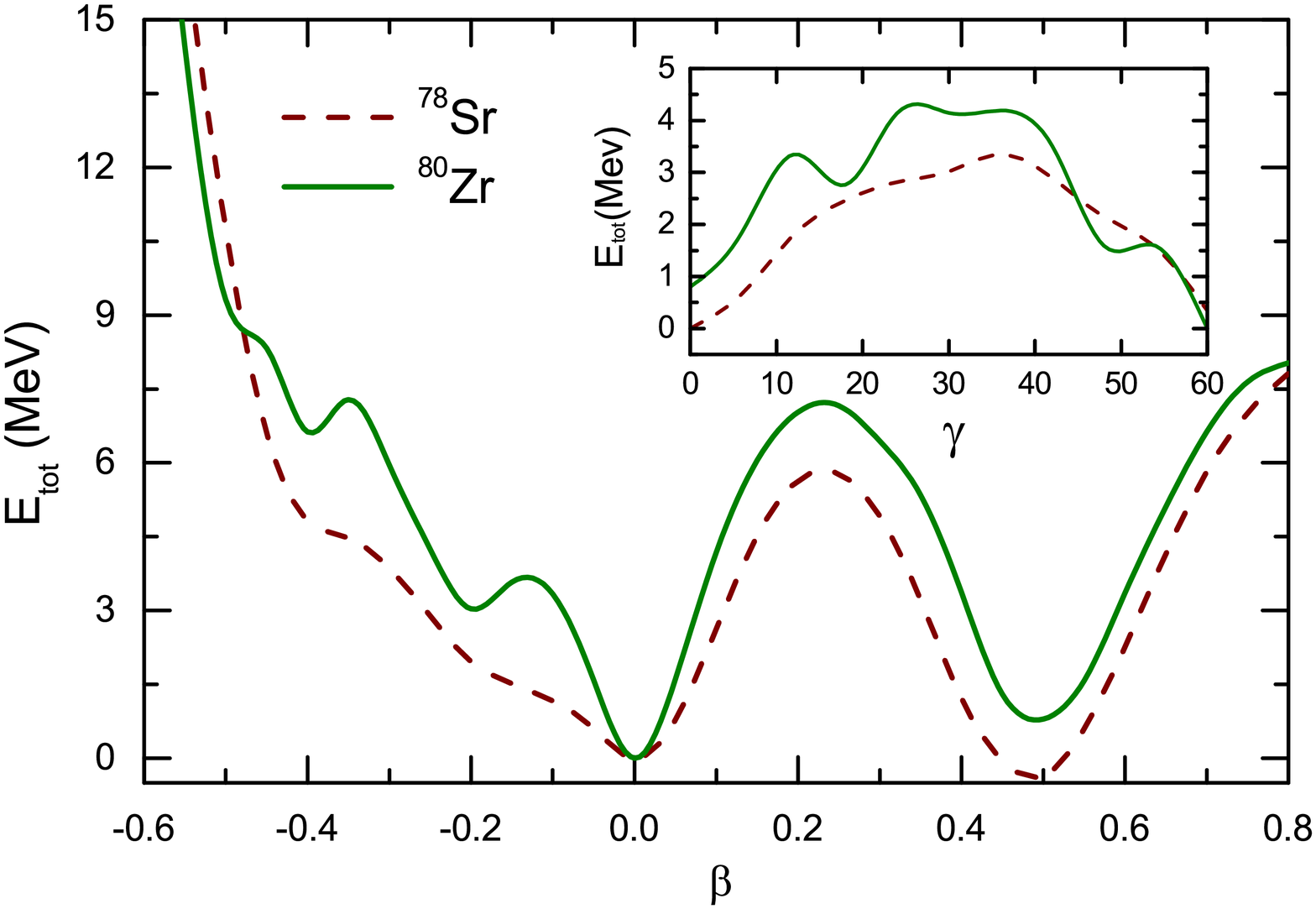}
\fi
\end{flushright}
\caption{(Color online) The total energies of  $^{78}$Sr and $^{80}$Zr as functions of axial deformation $\beta$. All energies are normalized to the spherical shape. The inset displays the PECs corresponding to the projections on the $\gamma$ deformation, namely the minimum for each $\gamma$ on the PES in the $\beta-\gamma$ plane.}
\label{fig:PEC}

\end{figure}

In this part we focus on the shape coexistence of the neutron-deficient $N=40$ isotones as shown in Fig.~\ref{fig:PES}. Fig.~\ref{fig:PEC} displays the potential energy curves (PECs) of $^{78}$Sr and $^{80}$Zr as functions of axial deformation parameter $\beta$.  The inset shows the PECs corresponding to the projections on the $\gamma$ deformation, corresponding to the minimum for each $\gamma$ on the PES in the $\beta-\gamma$ plane. In these two nuclei the coexisting spherical and prolate minima with very closed binding energies are observed, which are separated by certain barriers. The axial barriers of $^{78}$Sr and $^{80}$Zr are as high as $\sim$6~MeV and $\sim$8~MeV, respectively. After considering the $\gamma$ degree of freedom, the barrier heights are lowered down to 3.5~MeV and 4.5~MeV, respectively.

\begin{figure}[!htbp]
\begin{flushright}
\ifpdf
\includegraphics[width=0.75\textwidth]{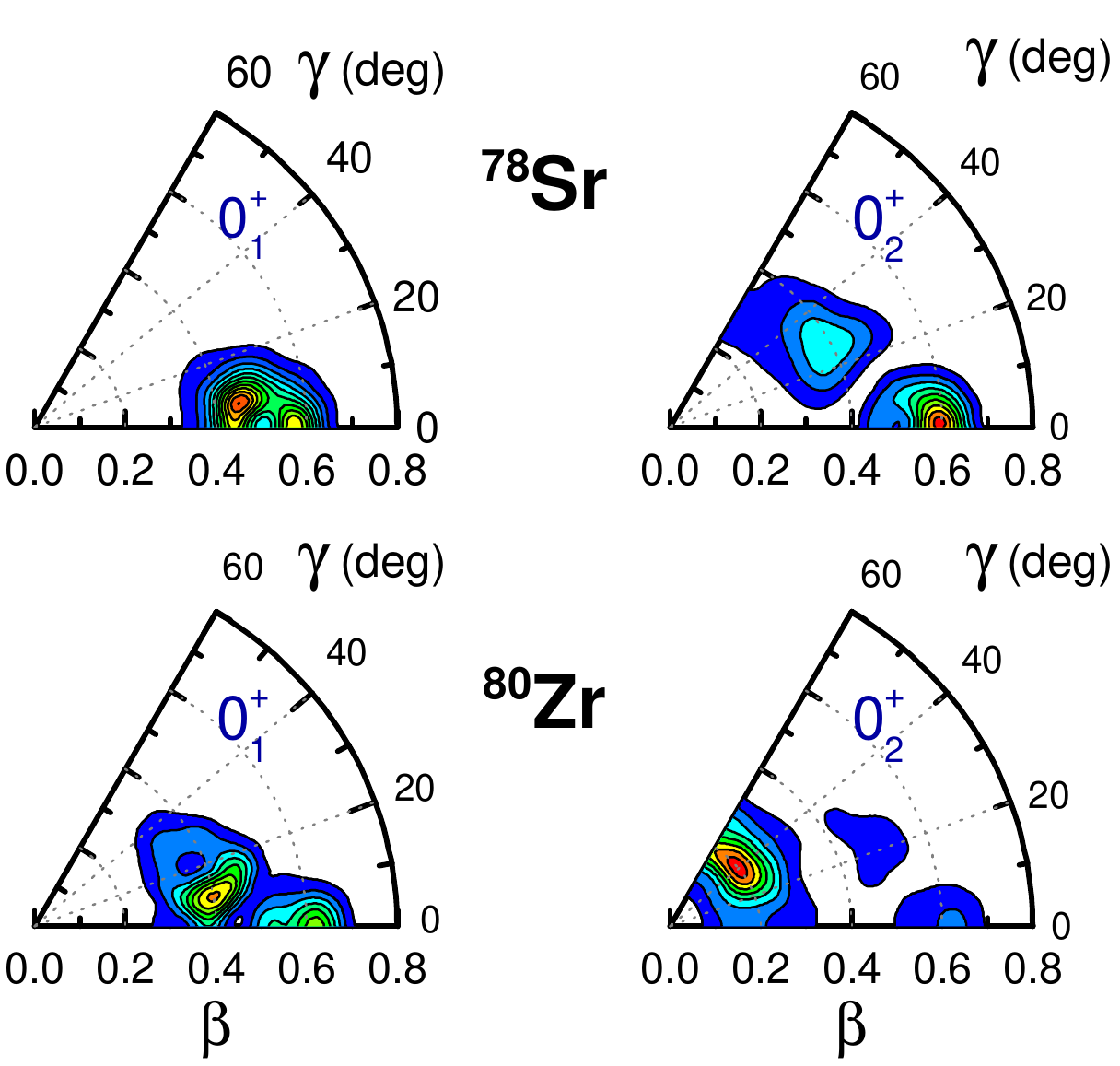}
\else
\includegraphics[width=0.75\textwidth]{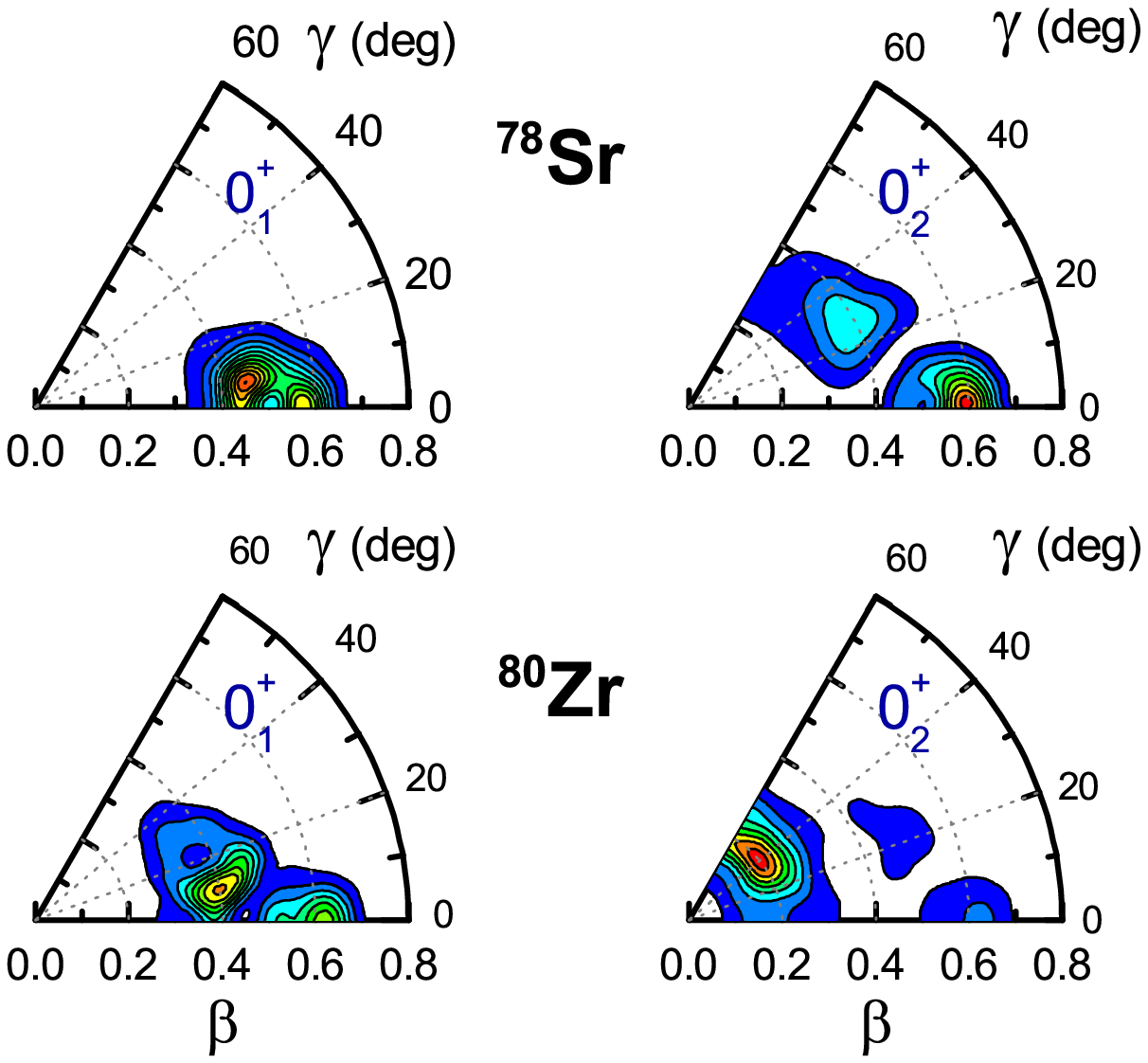}
\fi
\end{flushright}
\caption{(Color online) Distribution of the probability density $\rho_{J\alpha}(\beta,\gamma)$ for the first two $0^+$ states in $^{78}$Sr and $^{80}$Zr. See the text for details.}
\label{fig:wave}
\end{figure}

Shape mixing is an important quantum concept in the study of shape coexistence in atomic nuclei. Fig.~\ref{fig:wave} displays the distribution of probability density $\rho_{J\alpha}(\beta,\gamma)$  in the $\beta-\gamma$ plane for the first two $0^+$ states in $^{78}$Sr and $^{80}$Zr, where $\rho_{J\alpha}(\beta,\gamma)$ satisfies the following normalization condition:
\begin{equation}
  \int^\infty_0\beta d\beta\int^{2\pi}_0|\sin{3\gamma}|d\gamma\rho_{J\alpha}(\beta,\gamma)=1.
\end{equation}
The ground states $0_1^+$ of $^{78}$Sr and $^{80}$Zr are all located at large prolate-deformed region, close to the prolate minima in the PESs. For the second $0^+$ state in  $^{78}$Sr, a strong mixing of prolate and triaxial shapes is found, while in $^{80}$Zr the dominate configuration locates at the region with small oblate deformation. This is because $^{80}$Zr has a higher and wider triaxial barrier between two minima (c.f. Fig.~\ref{fig:PEC}).

\begin{figure}[!htbp]

\begin{flushright}
\ifpdf
\includegraphics[width=0.85\textwidth]{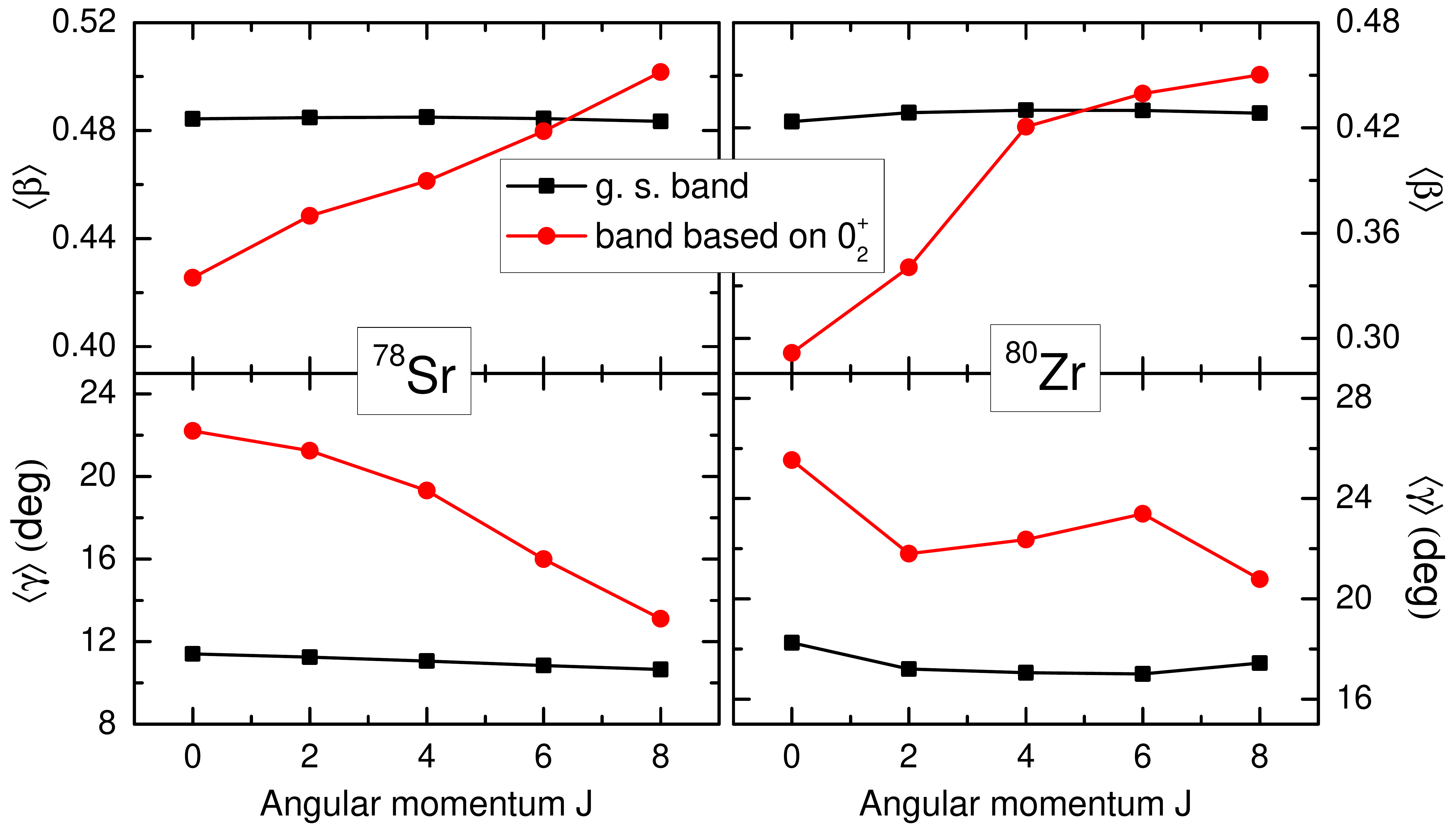}
\else
\includegraphics[width=0.85\textwidth]{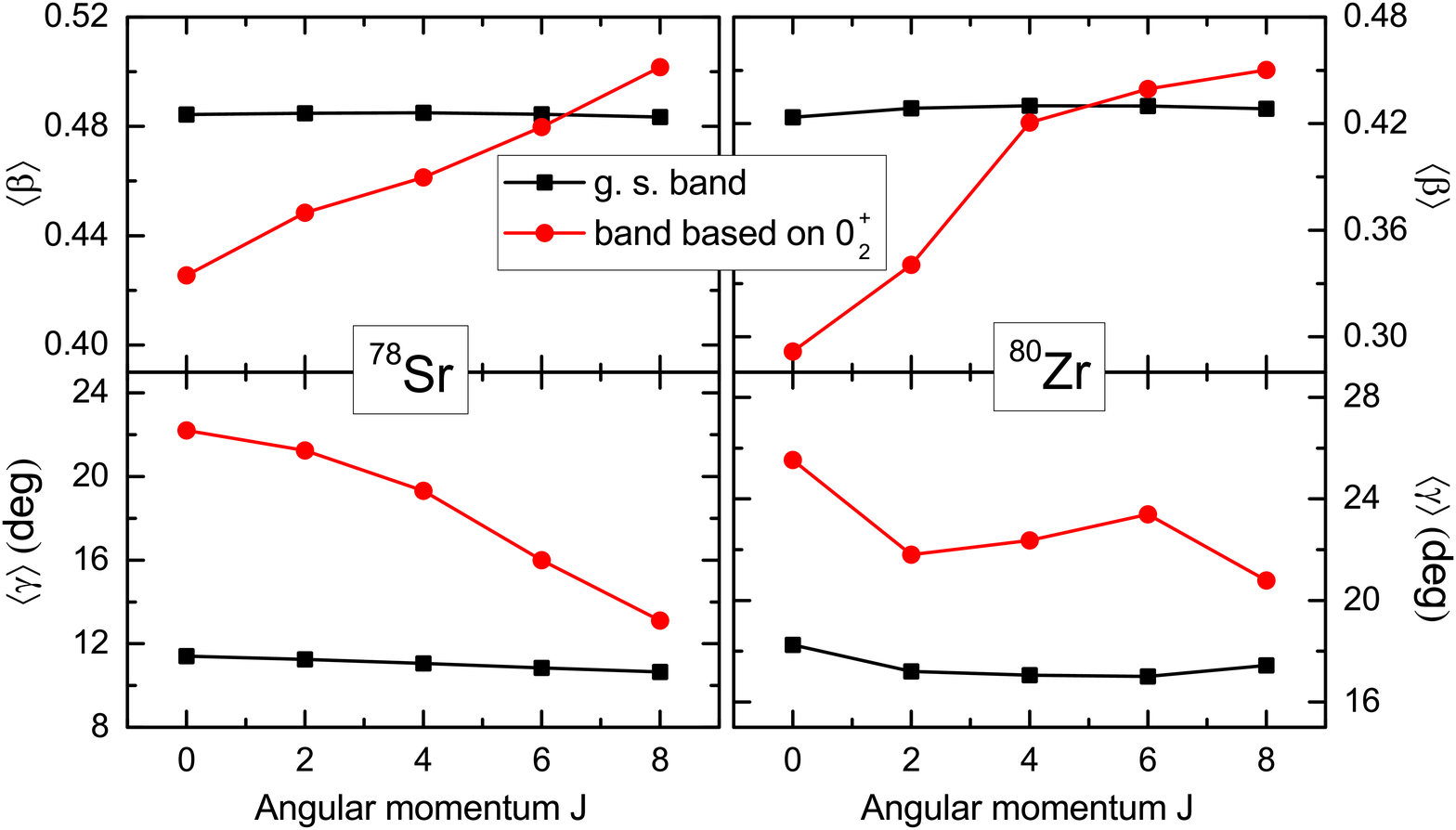}
\fi
\end{flushright}

\caption{(Color online) Average values $\langle\beta\rangle$ and $\langle\gamma\rangle$ as functions of angular momentum for the ground-state band and band  based on $0^+_2$ state in $^{78}$Sr (left panels) and $^{80}$Zr (right panels). See the text for details.}
\label{fig:bg}

\end{figure}

The effect of the barrier on the shape coexistence phenomenon can be further demonstrated by analyzing the evolution of the expected deformations $\left<\beta\right>$ and $\left<\gamma\right>$ of the states characterized by different shapes when increasing angular momentum. Fig.~\ref{fig:bg} shows the average values of $\langle\beta\rangle$ and $\langle\gamma\rangle$ as functions of angular momentum for the ground-state (g.s.) band and band  based on $0^+_2$ state in $^{78}$Sr and $^{80}$Zr. The $\langle\beta\rangle$ and $\langle\gamma\rangle$ of the ground state in $^{78}$Sr are $\sim0.48$ and $\sim11^o$, respectively, and keep almost constant as the angular momentum increases. Similar systematics are found in $^{80}$Zr, but with a smaller $\langle\beta\rangle$ and larger $\langle\gamma\rangle$. For the second $0^+$ state in $^{78}$Sr, the $\langle\beta\rangle$ and $\langle\gamma\rangle$  are $\sim0.42$ and $22^o$, respectively, and approach to the expectations of the ground-state band gradually when increasing angular momentum. This means that the higher-spin state in the excited band has located beyond the barrier and can 'talk' easily with the ground-state band. However, in $^{80}$Zr the $\langle\gamma\rangle$ of the excited band is larger than $20^o$ and keeps away from the ground-state band, even though the  $\langle\beta\rangle$ approaches to the ground-state band rapidly. This can be attributed to the higher and wider triaxial barrier in $^{80}$Zr.

\begin{figure}[!htbp]
\begin{flushright}
\ifpdf
\includegraphics[width=0.9\textwidth]{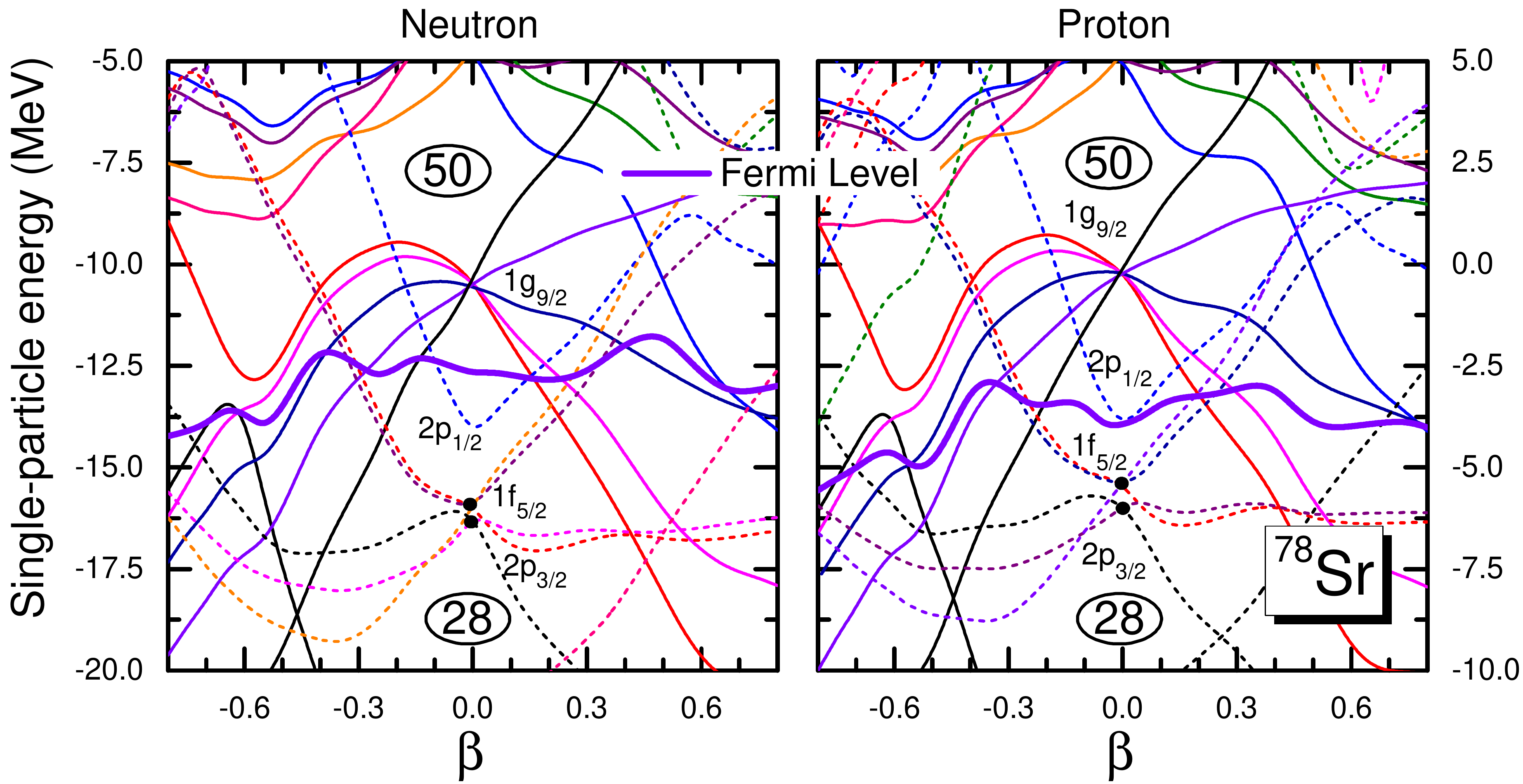}
\else
\includegraphics[width=0.9\textwidth]{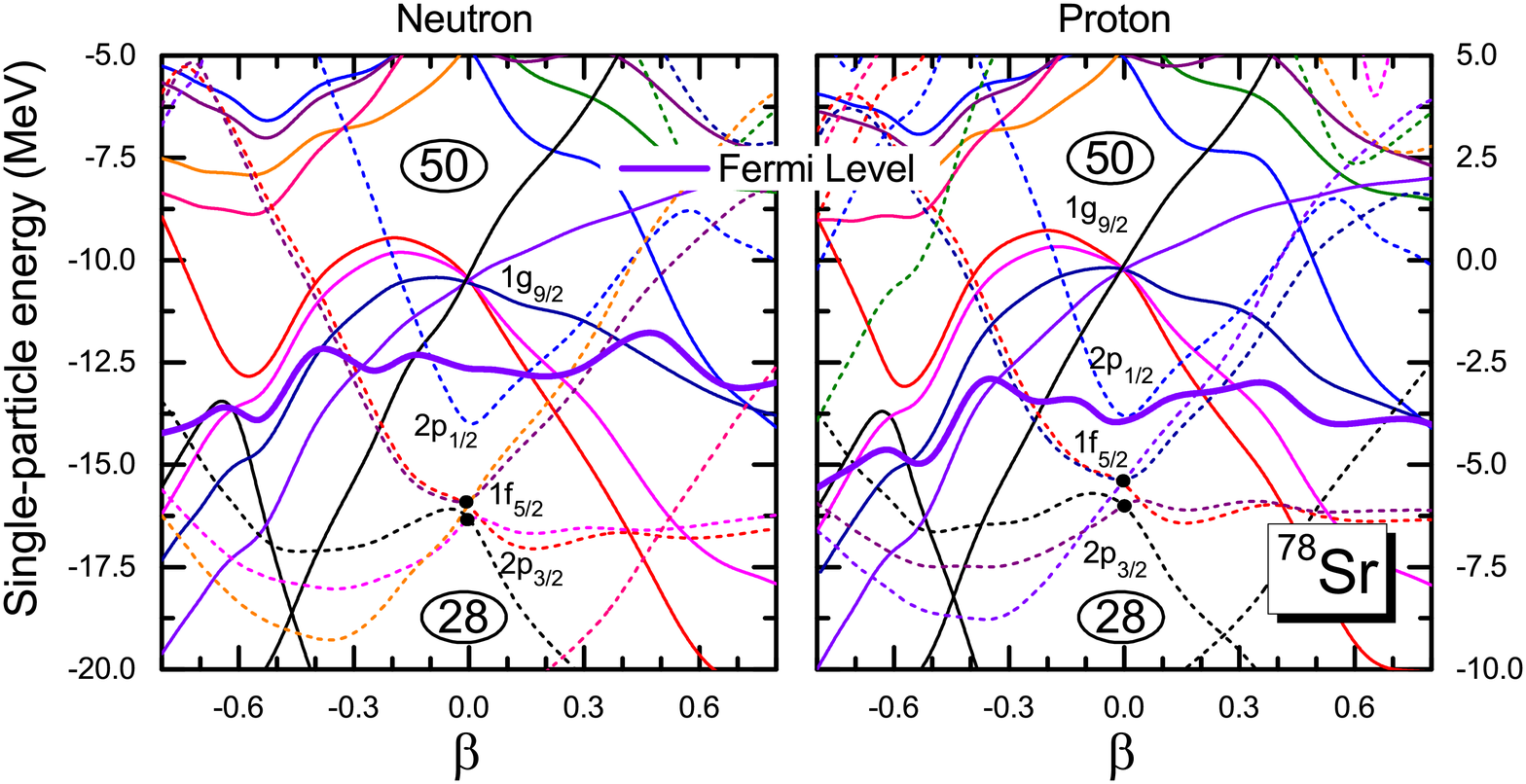}
\fi
\end{flushright}
\caption{(Color online) Neutron and proton single-particle levels for $^{78}$Sr as functions of the axial deformation parameter $\beta$. The ultra thick purple lines denote the Fermi levels. See the text for details.}
\label{fig:EPS}

\end{figure}

A microscopic picture of the coexisting minima in PECs and related phenomenon of the low-lying states in neutron-deficient $N=40$ isotones emerges when considering the single-particle levels. Taking $^{78}$Sr as an example, we plot the single-particle levels  as functions of the axial deformation parameter $\beta$ in Fig.~\ref{fig:EPS}. It is shown that the neutron Fermi level goes across the low-level-density region at ${\beta\approx0}$, giving rise to the spherical minimum. While the proton Fermi level locates in the middle of the energy gap  around ${\beta\approx0.5}$,  which gives rise to the prolate minimum in $^{78}$Sr.

\subsection{Spectroscopy of low-lying states in $N=40$ isotones with $Z>28$}

\begin{figure}[!htbp]
\begin{flushright}
\ifpdf
\includegraphics[width=0.49\textwidth]{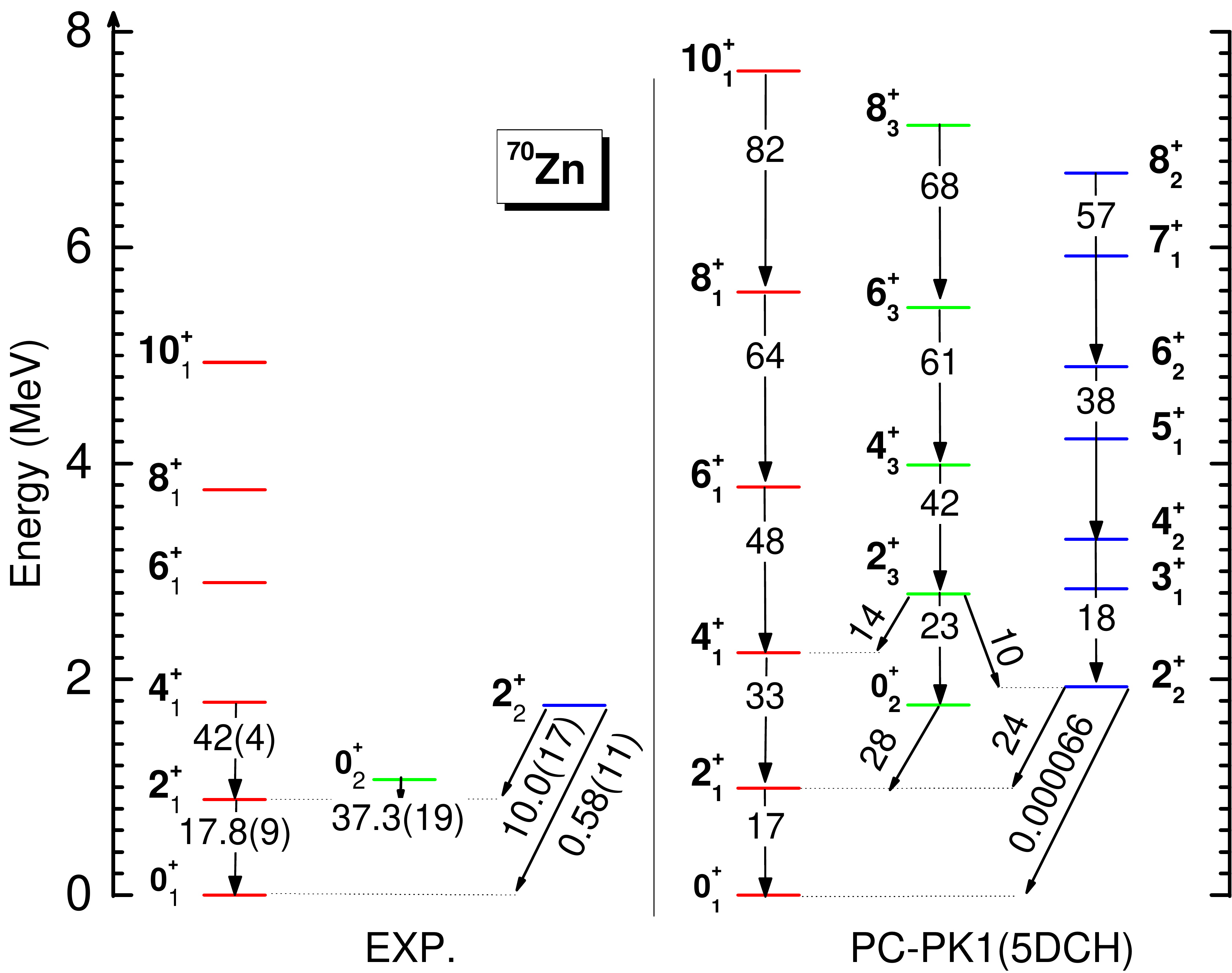}\hfill\includegraphics[width=0.49\textwidth]{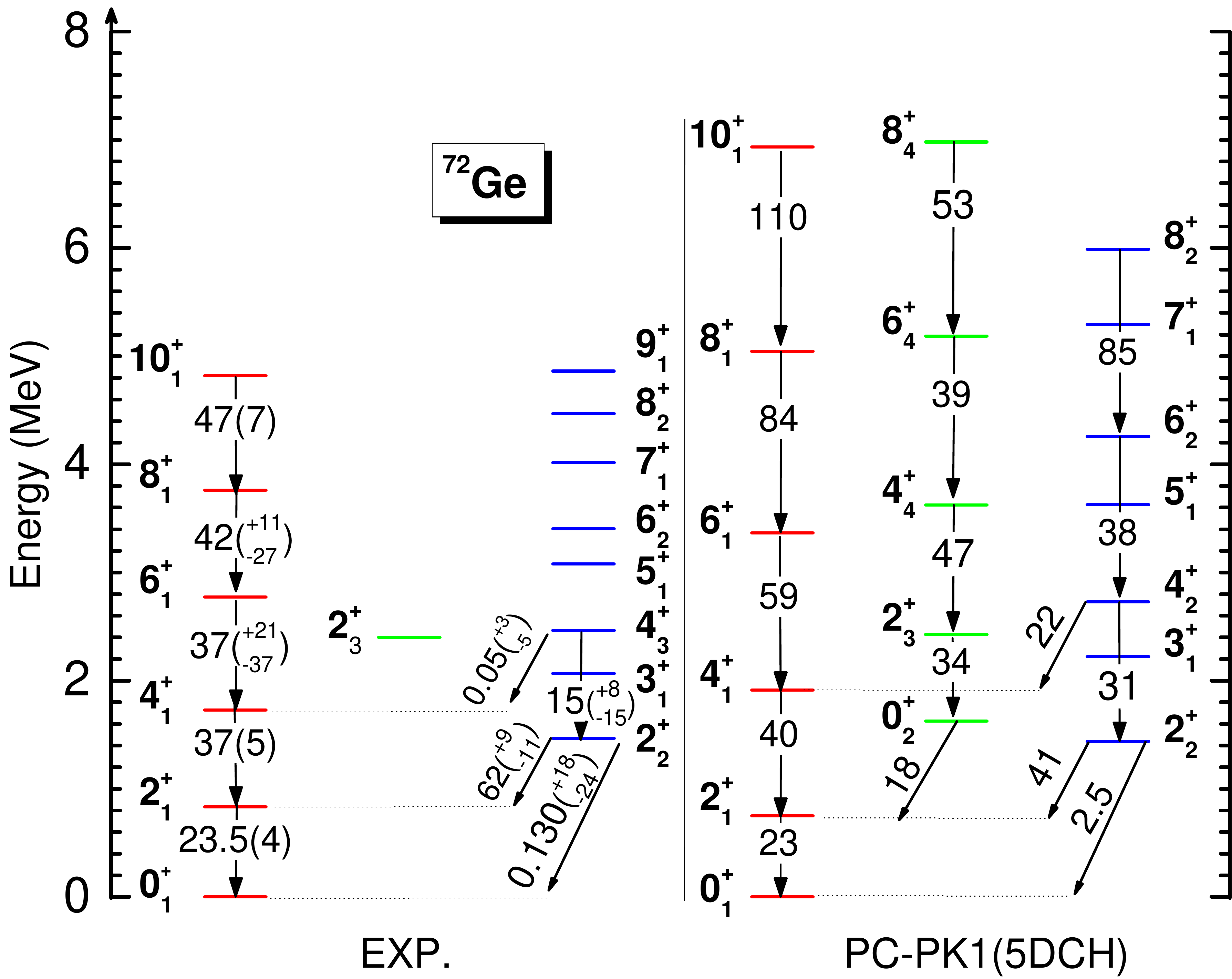}
\else
\includegraphics[width=0.49\textwidth]{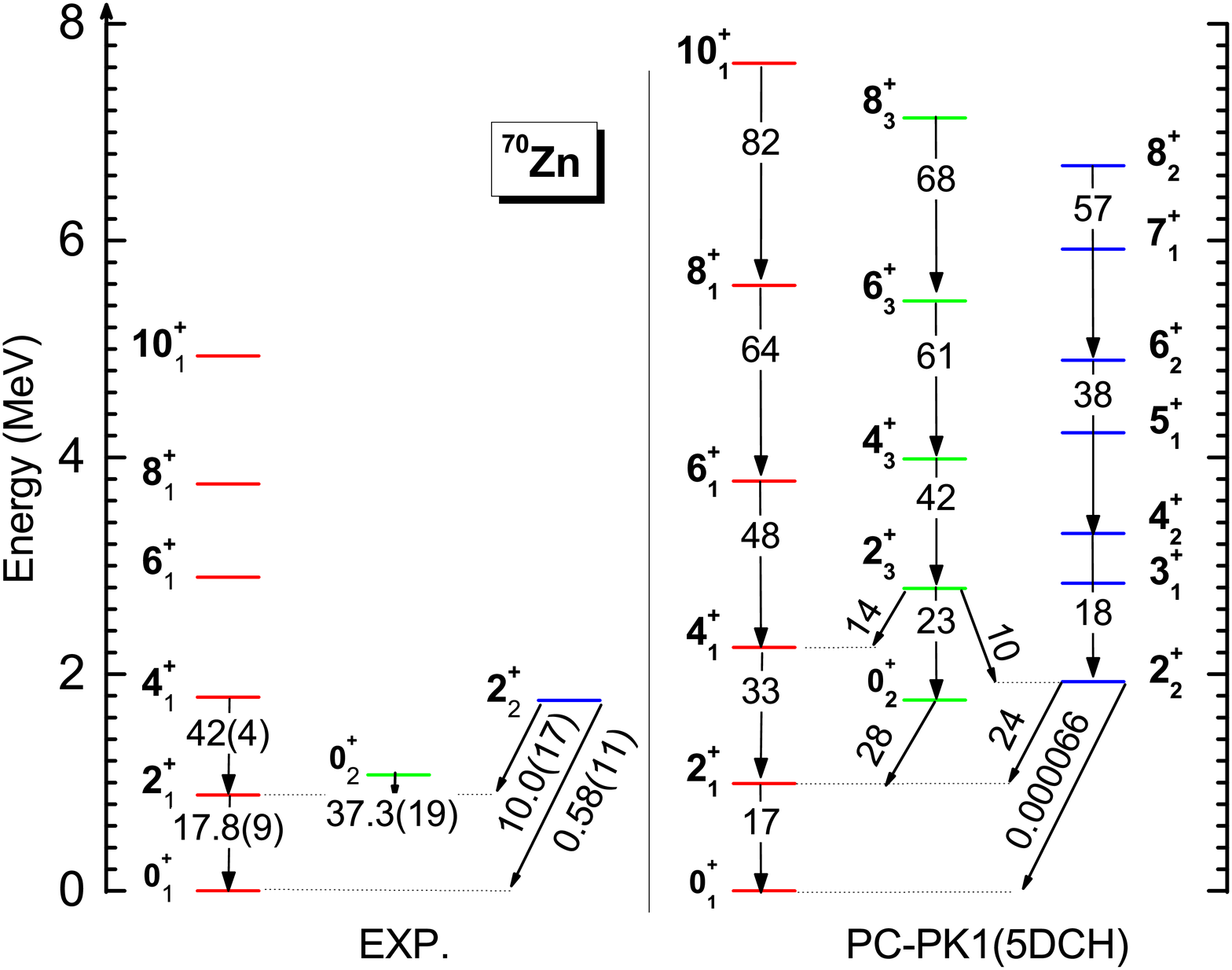}\hfill\includegraphics[width=0.49\textwidth]{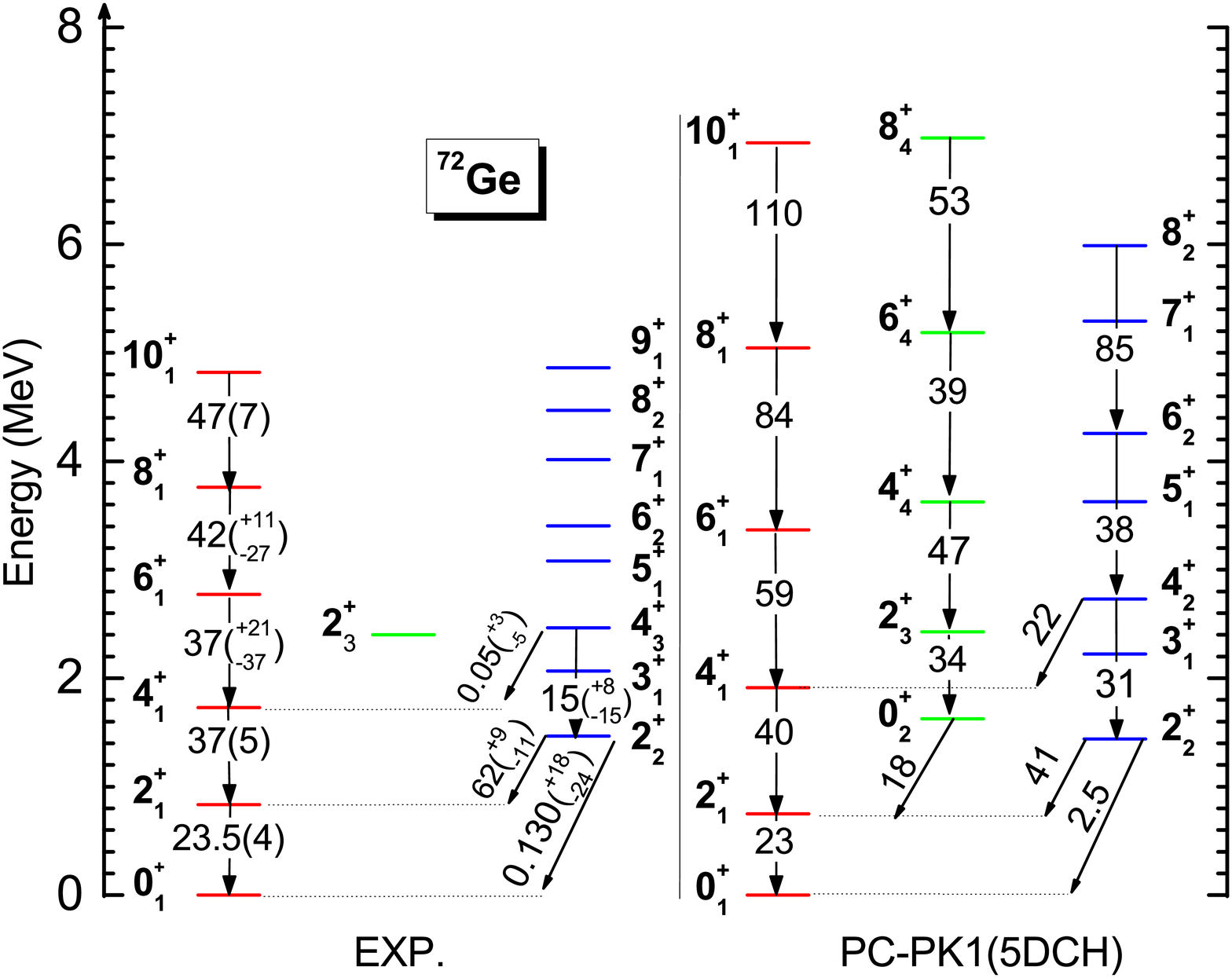}
\fi
\end{flushright}
\caption{(Color online) Low-lying spectra of $^{70}$Zn (left plot) and $^{72}$Ge (right plot) in comparison with the available data. The $B(E2)$ transition strengths are given in Weisskopf units (W.u.). The experimental data are taken from Refs.~\cite{Brookhaven,Berkeley,PhysRevC.79.054310,Sun2014} }
\label{fig:spectra-Zn70}

\end{figure}
%
%

\begin{figure}[!htbp]
\ifpdf
\includegraphics[width=0.49\textwidth]{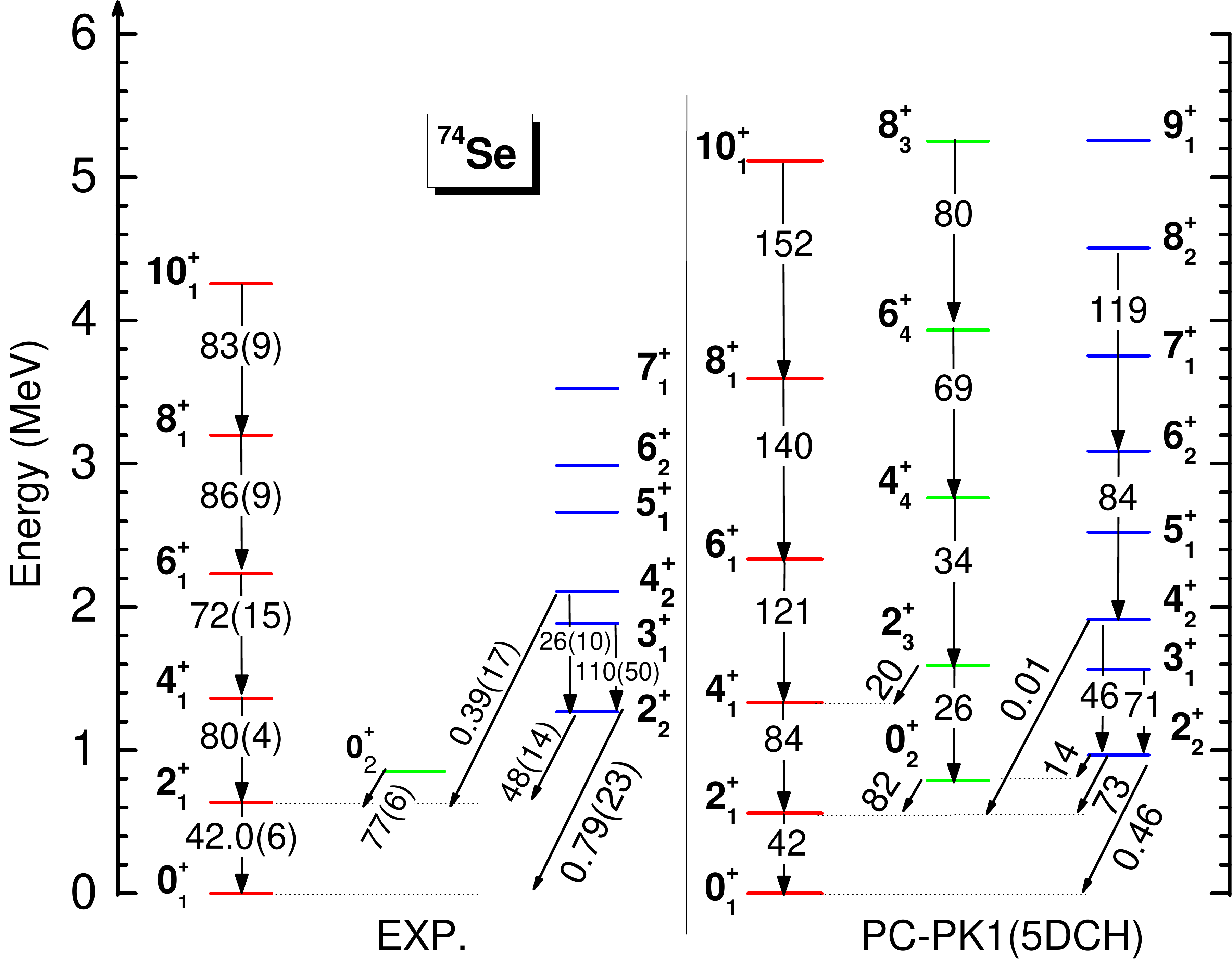}\hfill\includegraphics[width=0.49\textwidth]{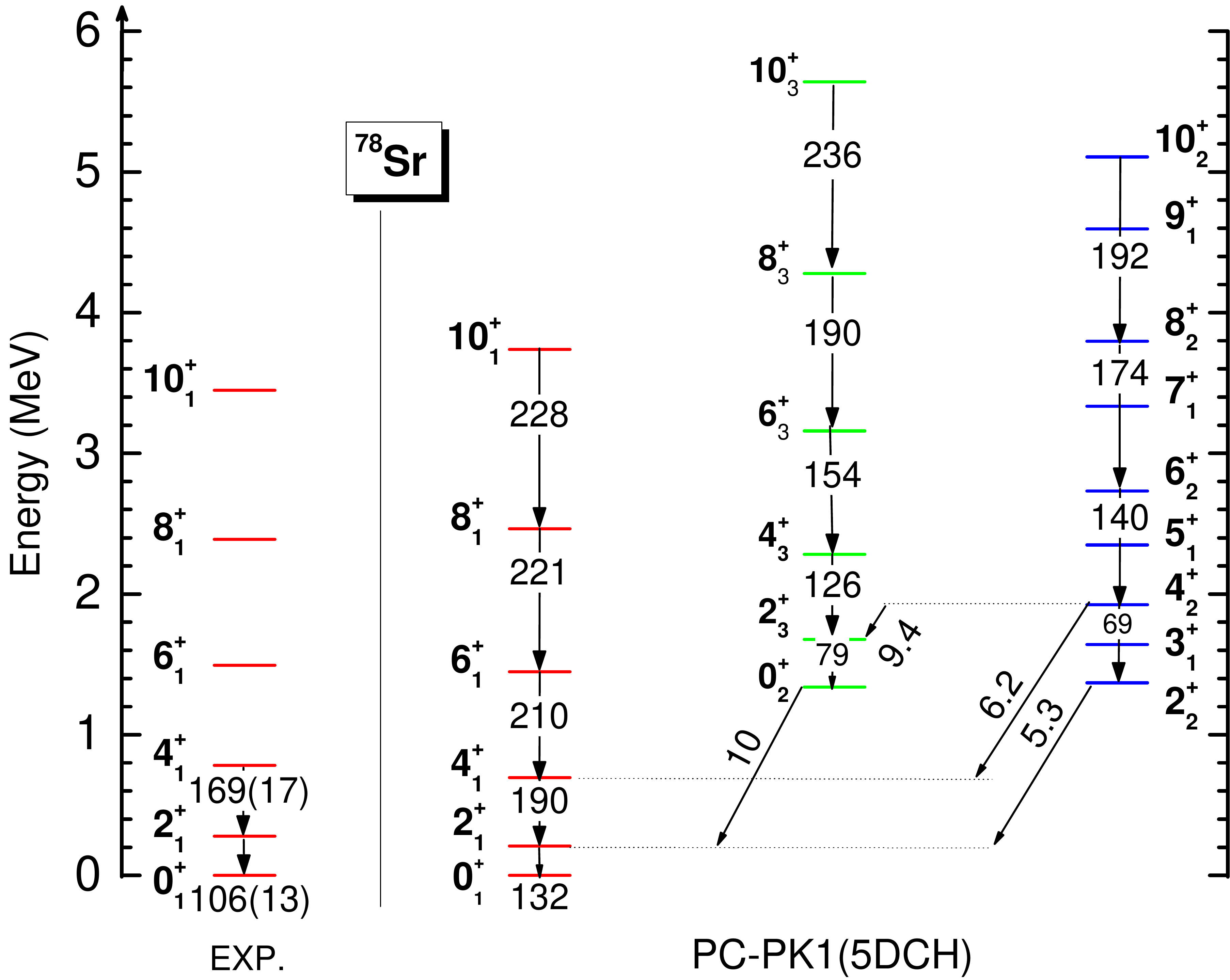}
\else
\includegraphics[width=0.49\textwidth]{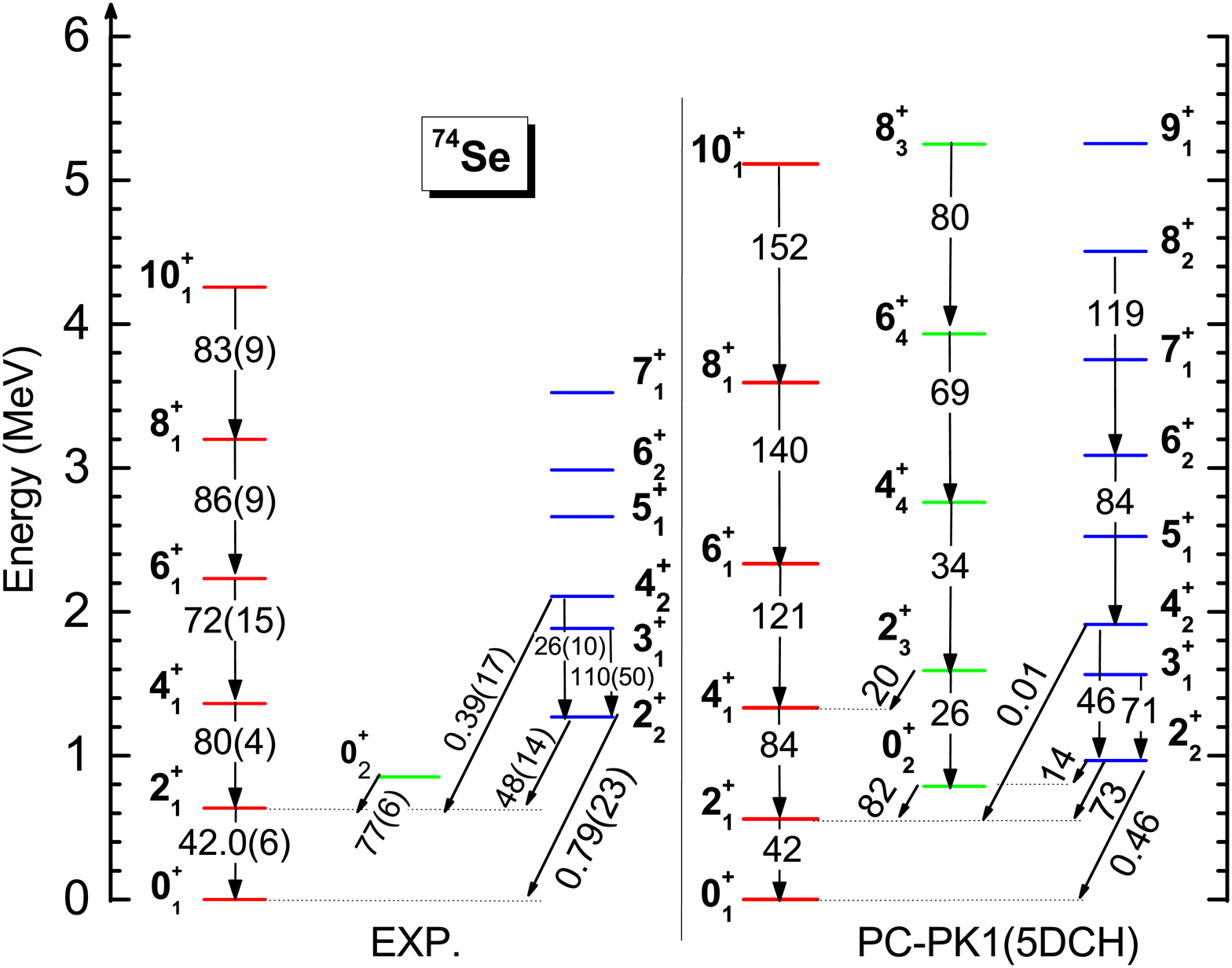}\hfill\includegraphics[width=0.49\textwidth]{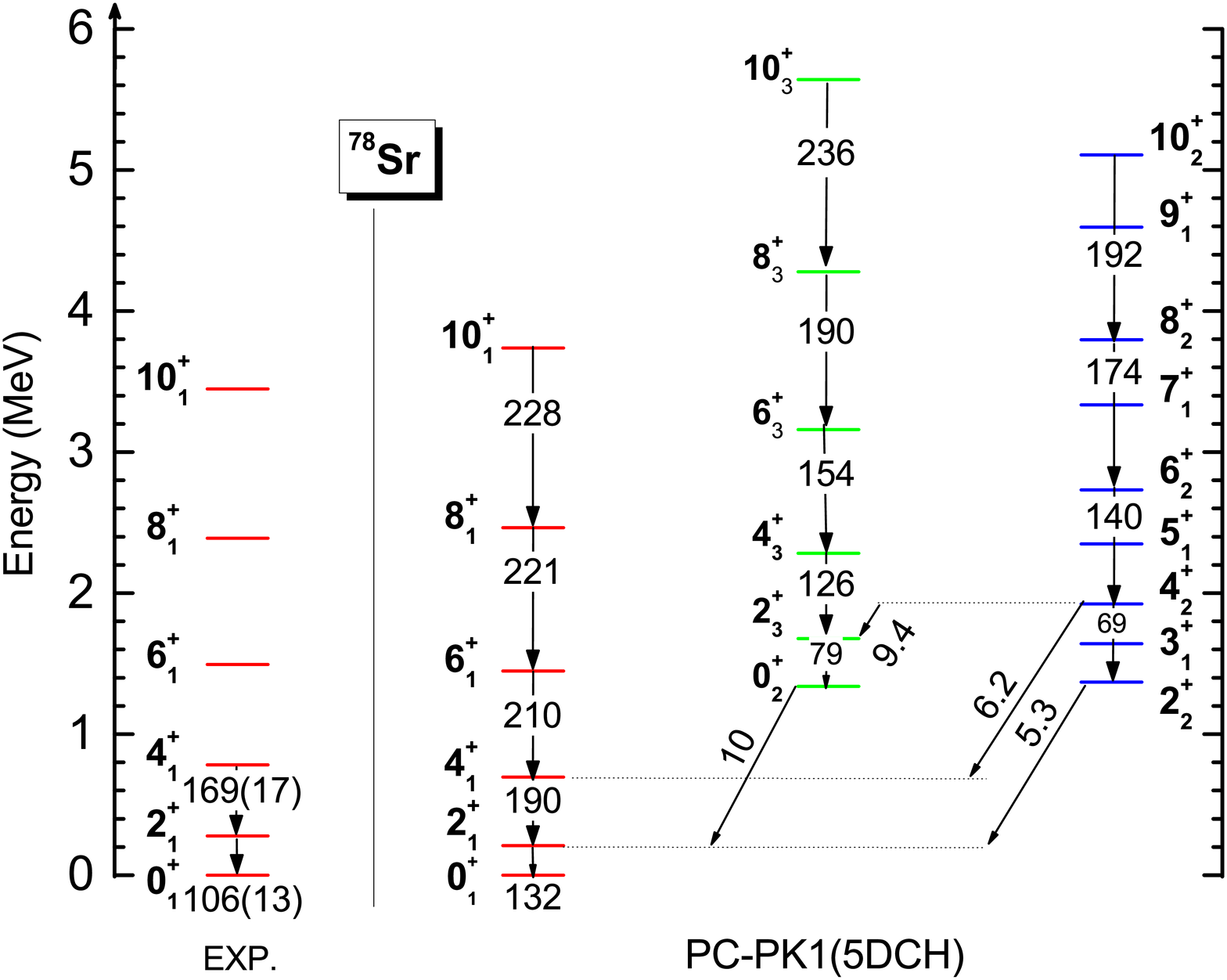}
\fi
\caption{(Color online) Same as Fig.~\ref{fig:spectra-Zn70}, but for $^{74}$Se (left plot) and $^{78}$Sr (right plot), and the experimental data are taken from Refs.~\cite{Brookhaven,Berkeley}.  }
\label{fig:spectra-Se74}
\end{figure}


%

\begin{figure}[htbp]

\begin{flushright}
\ifpdf
\includegraphics[width=0.49\textwidth]{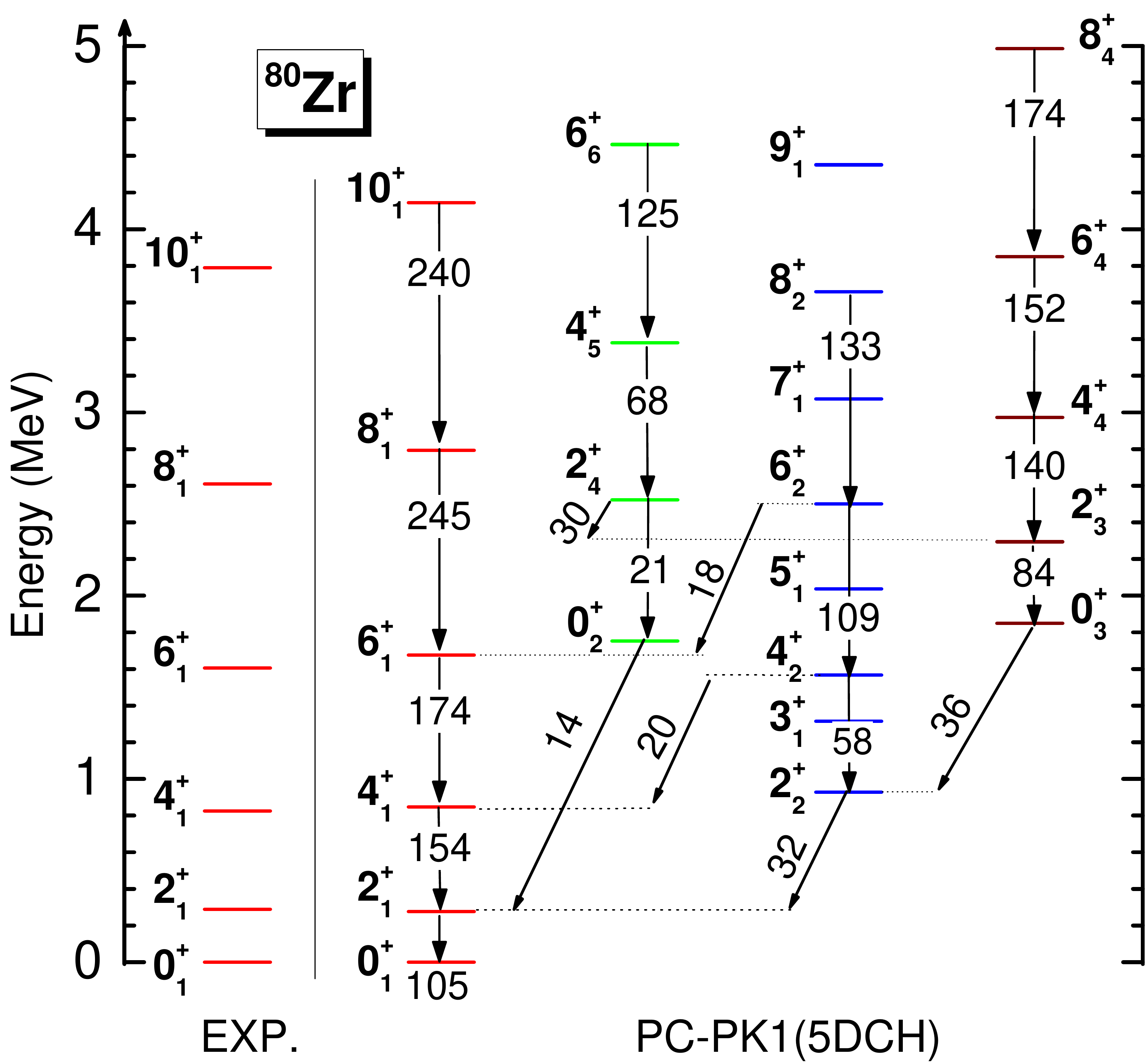}
\else
\includegraphics[width=0.49\textwidth]{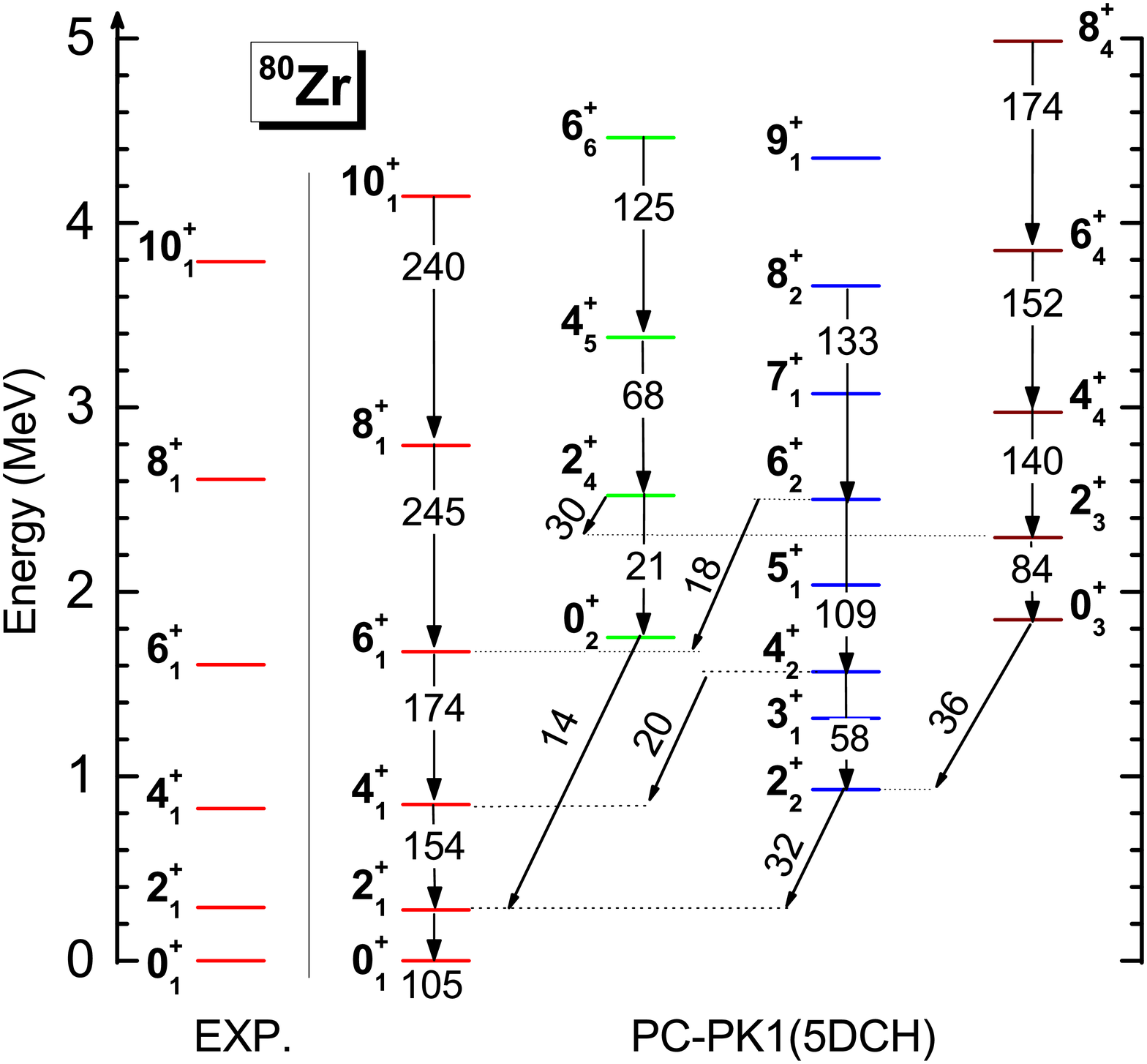}
\fi
\end{flushright}
\caption{(Color online) Same as Fig.~\ref{fig:spectra-Zn70}, but for $^{80}$Zr, and the experimental data are taken from Ref.~\cite{Brookhaven}. }
\label{fig:spectra-Zr80}

\end{figure}

The calculated low-lying spectra  of $N=40$ isotones with $Z>28$ are compared with experimental data in Figs.~\ref{fig:spectra-Zn70} --~\ref{fig:spectra-Zr80}. Since the one of $^{76}$Kr has been discussed in detail in Refs.~\cite{Fu2013,Yao2014}, here we do not repeat it again. One can see that the ground state bands of these nuclei are reproduced well except those with $I\geq4$ in $^{70}$Zn and $^{72}$Ge, which are too spread compared to the data. As can be seen in Figs.~\ref{fig:spectra-Zn70}, \ref{fig:spectra-Se74} and Fig.~6 in Ref.~\cite{Fu2013}, the excitation energies of quasi-$\beta$ and quasi-$\gamma$ bands in $^{70}$Zn, $^{74}$Se, and $^{76}$Kr are reproduced quite well by 5DCH calculation.  Both the calculated intraband and interband transitions are in good agreement with the available data concerning that our model is completely parameter free. Although it is not straightforward to obtain the evident proof from Figs. \ref{fig:spectra-Zn70}-\ref{fig:spectra-Zr80} for the shape coexistence in these nuclei, the well reproduced low-lying spectra, on the other hand, illustrate the predictive reliability of our model.

\section{Summary and Perspectives \label{summary}}

In summary, we have presented a systematic beyond RMF study of the low-lying states in the even-even $N=40$ isotones. The excitation energies and electric transition strengths have been obtained by solving a five-dimensional collective Hamiltonian with the collective parameters determined from the RMF calculations using the PC-PK1 force. The results have been compared with experimental data and those obtained by similar calculations but using the Gogny force D1S. The theoretical calculations can reproduce not only the systematics of the low-lying states along the isotonic chain but also the detailed structure of the spectroscopy in singular nucleus. We find a picture of spherical-oblate-prolate shape transition in the $N=40$ isotones. The coexistence of low-lying excited $0^+$ states has also been shown to be a common feature in neutron-deficient $N=40$ isotones.

The present study provides an example demonstrating how CDFT can describe not only nuclei with a single-configuration-dominated structure but also nuclei with a coexistence structure of distinctly different shapes in low-excitation energy and the transition behavior. Although the commonly used CDFT without tensor coupling term can describe most of the observalbes in the $N=40$ isotones but fail to reproduce the onset of the deformation of ground state in neutron-rich $N=40$ isotones. Therefore, the 5DCH calculation  based on the deformed DDRHF+BCS with PKA1 effective interaction, where the tensor coupling of of $\pi$ and $\rho$ mesons is included self-consistently, is highly demanded. Work along this direction is in progress.

\ack
This work was supported in part by the National Natural Science Foundation of China under Grants Nos. 11375076, 11475140 and 11105110, the Natural Science Foundation of Chongqing under Grant No. cstc2011jjA0376, the Fundamental Research Funds for the Central Universities under Grant No. XDJK2011B002, and the Specialized Research Fund for the Doctoral Program of Higher Education under Grant No. 20130211110005.

%
\providecommand{\newblock}{}

\end{document}